**Isospin magnetic texture and intervalley exchange interaction in rhombohedral tetralayer graphene**

Nadav Auerbach[1†], Surajit Dutta[1†], Matan Uzan[1†], Yaar Vituri[1], Yaozhang Zhou[1], Alexander Y. Meltzer[1], Sameer Grover[1], Tobias Holder[2], Peleg Emanuel[1], Martin E. Huber[3], Yuri Myasoedov[1], Kenji Watanabe[4], Takashi Taniguchi[5], Yuval Oreg[1], Erez Berg[1], and Eli Zeldov[1]*

The tunable band structure and nontrivial topology of multilayer rhombohedral graphene lead to a variety of correlated electronic states with isospin orders – meaning ordered states in the combined spin and valley degrees of freedom – dictated by the interplay of spin-orbit coupling and Hund's exchange interactions. However, methods for mapping local isospin textures and determining the exchange energies are currently lacking. Here, we image the magnetization textures in tetralayer rhombohedral graphene using a nanoscale superconducting quantum interference device. We observe sharp magnetic phase transitions that indicate spontaneous time-reversal symmetry breaking. In the quarter-metal phase, the spin and orbital moments align closely, providing a bound on the spin-orbit coupling energy. We also show that the half-metal phase has a very small magnetic anisotropy, which provides an experimental lower bound on the intervalley Hund's exchange interaction energy. This is found to be close to its theoretical upper bound. The ability to resolve the local isospin texture and the different interaction energies will allow a better understanding of the phase transition hierarchy and the numerous correlated electronic states arising from spontaneous and induced isospin symmetry breaking in graphene heterostructures.

[1]Department of Condensed Matter Physics, Weizmann Institute of Science, Rehovot 7610001, Israel
[2]School of Physics and Astronomy, Tel Aviv University, Tel Aviv, Israel.
[3]Departments of Physics and Electrical Engineering, University of Colorado Denver; Denver, Colorado 80217, USA
[4]Research Center for Electronic and Optical Materials, National Institute for Materials Science; 1-1 Namiki, Tsukuba 305-0044, Japan
[5]Research Center for Materials Nanoarchitectonics, National Institute for Materials Science; 1-1 Namiki, Tsukuba 305-0044, Japan
[†]These authors contributed equally to this work
*eli.zeldov@weizmann.ac.il

Magnetism has been studied since antiquity, yet understanding how magnetic order emerges in complex, interacting systems remains a challenge. In solids, magnetism generally stems from electron spin or orbital motion. While spin moments dominate in most materials, recent studies on two-dimensional (2D) van der Waals systems reveal a leading role for orbital magnetism. Unlike spin-based magnetism, orbital magnetism reflects the distribution of electron wavefunctions and encodes the band topology and Berry curvature. In addition to spin, electrons in graphene-like 2D materials exhibit a valley degree of freedom, which strongly influences their behavior. Crucially, spontaneous magnetization is forbidden if time-reversal or full spin-valley isospin symmetry is preserved. Thus, magnetism in these systems offers key insights into symmetry breaking and electronic interactions in strongly interacting 2D materials [1–3].

An important characteristic of atomic-layer materials is their highly tunable electronic structure, which, together with a high density of states (DOS) and nontrivial band geometry, can drive spontaneous symmetry breaking and formation of magnetic ground states even at zero magnetic field [4–8]. The resulting isospin symmetry breaking produces a rich array of textures, including spin and valley polarization, inter-valley coherence (IVC), spin-valley locking via spin-orbit coupling (SOC), and skyrmion-like patterns [9–14]. Rhombohedral multilayer graphene (RMG) has recently emerged as a low-disorder platform for realizing such phase [15–30], including correlated insulators [21,22,30,31], Chern insulators [31,32], ferromagnets [15–19,21,28,31–36], ferroelectrics [28], superconductors [23,35,37–43], and multiferroics [16]. Furthermore, proximity-induced SOC can stabilize certain phases [33,34,37–43], while moiré engineering enables exotic fractional quantum states [15,18,19,32,44,45].

In the absence of interactions, electronic bands with different valley and spin polarizations are degenerate. A key factor determining the energetic hierarchy of isospin symmetry-broken states is the Hund's exchange interaction, which describes interactions between electrons in the different isospin bands, comprising intravalley and intervalley components. While intravalley exchange between opposite spins is usually the dominant energy due to the long-range nature of Coulomb interactions, intervalley exchange crucially shapes the material's low-energy physics. For example, in the absence of SOC, intervalley Hund's coupling favors a spin-ferromagnetic state over spin-antiferromagnetic states. SOC, in contrast, tends to stabilize intervalley spin-antiferromagnetic order in the absence of Hund's coupling. Thus, observing ferromagnetism indicates dominant intervalley Hund's interaction. It also permits spin-ferromagnetic IVC (charge density wave), in contrast to antiferromagnetic IVC (spin density wave). Despite its fundamental significance, a direct measurement of the Hund's interaction energy $U_{Hu}$ has been lacking. In RMG, $U_{Hu}$ is believed to be ferromagnetic, but directly probing its strength is challenging because it requires a perturbation that couples oppositely to spins in different valleys— unavailable experimentally. However, intrinsic Ising SOC ($\lambda_{SOC}$) acts exactly as such a perturbation, generating a small spin anisotropy of the order of $\lambda_{SOC}^2/U_{Hu}$. By rotating a small external magnetic field, we probe this anisotropy, enabling the quantitative measurement of $U_{Hu}$.

Previous studies on isospin ordering in van der Waals materials have largely relied on global measurements and theoretical models [15–19,31–34,37,38,46]. Here, we introduce a nanoscale magnetic imaging technique to directly visualize local isospin textures in symmetry-broken phases of crystalline ABCA graphene. We further explore how magnetic fields affect isospin state stability, particularly via interactions with Berry curvature and orbital magnetization. The field influences both spin moments through the Zeeman effect and orbital moments via a valley-Zeeman effect, resulting in band shifts [20,47]. Our results show that the combination of momentum-dependent Berry curvature and unique band structure induces nontrivial magnetic band reconstruction, establishing RMG as a versatile platform for exploring the interplay of topology and interaction-driven physics [12].

### Transport and topological magnetic band reconstruction

Three dual-gated ABCA graphene devices were studied, showing qualitatively similar behavior (Fig. 1 and Extended Data Figs. 1,5). The *dc* top and bottom gates voltages $V_{tg}^{dc}$ and $V_{bg}^{dc}$ control carrier density $n$ and out-of-plane displacement field $D$. Figure 1d presents the longitudinal resistance $R_{xx}(n, D)$ at zero out-of-plane magnetic field ($B_z = 0$) in device A, which displays metallic behavior with $R_{xx} < 1$ kΩ for $|n| \gtrsim$ 0.5×10$^{12}$ cm$^{-2}$. Several ridge-like features partition the phase diagram. We focus on the weak ridge, $n_a(D)$ (white dashed line, Fig. 1d), which traces a kink in the quantum oscillations (QOs) when $B_z$ is raised to 3 T (Fig. 1e). To its left, vertical QOs reflect a simple band structure with a single Fermi surface (left inset, Fig. 1c) and four-fold degenerate Landau levels (LLs). Our single particle band structure calculations (Extended Data Fig. 3), accurately reproduce both the transition line $n_a(D)$ and the QOs in regions A and B (Fig. 1f). The $n_a(D)$ line reflects a Lifshitz transition—annulus opening in the valence band—producing a sharp step in DOS (black arrow, Fig. 1c). Since at $|n| > |n_a(D)|$ the band structure is simple and no significant interaction effects are expected due to low DOS, no additional transitions at higher densities are anticipated, as confirmed at zero field (Fig. 1d). Unexpectedly, in this full metal (M) phase, the QOs exhibit a break along the cyan dashed line $n_\pi(D)$ in Fig. 1e, across which the QOs display a $\pi$-shift. Such a shift has been observed previously in rhombohedral structures but unexplained [11]. The high-resolution measurement and low disorder allow resolving the $\pi$-shift as splitting of four-fold LLs and their recombination with neighboring LLs. As described in Methods, this is a result of a topological magnetic band reconstruction (TMBR) due to large Berry-curvature-induced orbital magnetization $\mathcal{M}_{SR}$ in the annular Fermi surface (Fig. 1a), which in the presence of $B_z$, gives rise to pronounced splitting in the DOS between $K$ and $K'$ valleys (Fig. 1c) and in band distortion, $\varepsilon(k) = \varepsilon_0(k) - \mathcal{M}_{SR}(k)B_z$ (Fig. 1b).

### Spontaneous symmetry breaking

The rich QOs in $R_{xx}$ in Fig. 1e reveal several symmetry broken states. The vertical QO lines in region A correspond to four-fold degenerate LLs in a full metal with single Fermi surface. The annulus opening at $n_a(D)$ introduces two Fermi surfaces in region B, with LLs from each. The TMBR further lifts valley degeneracy yielding a full metal with four sets of intersecting LLs (Extended Data Fig. 3). At lower $|n|$, interaction-driven isospin symmetry broking emerges. Regions D and F exhibit vertical QOs like region A, but with periods, $\Delta n = \frac{Nh}{eB_z}$, corresponding to LL degeneracies $N = 2$ and 1, revealing half-metal (1/2M) and quarter-metal (1/4M) phases with simple Fermi surfaces ($h$, Planck constant; $e$, elementary charge; see Extended Data Fig. 4).

### Imaging isospin magnetism

To probe the nature of the symmetry-broken states, we performed nanoscale magnetic imaging on device B utilizing a superconducting quantum interference device on the apex of a sharp pipette (SQUID-on-tip, SOT) [48,49]. A small *ac* voltage $V_{tg}^{ac}$ applied to the top gate modulates $n$ by $n_{ac}$ (Fig. 2d), inducing a differential change in the local magnetization $m(x, y)$ and corresponding stray field $B_z^{ac}(x, y)$, detected by the SOT. Figure 2a shows $B_z^{ac}(n, D)$ at a fixed location above the sample at a low $B_z = 10$ mT. Two sharp peak lines indicated abrupt increase in the magnetization $\mathcal{M}$ at M-to-1/2M and 1/2M-to-1/4M phase transitions, consistent with transport data (Extended Data Fig. 5). In a state that preserves time-reversal symmetry (TRS), the net magnetization vanishes. TRS breaking induces magnetization via charge imbalance in spin or valley sectors. The 2D graphene structure dictates a highly anisotropic orbital magnetic moment (schematic current loops in Figs. 2f-h) with an easy $\hat{z}$ axis. In contrast, our understanding of the spin anisotropy is lacking due to limited ability for direct experimental visualization of the spin properties. Contrary to the orbital moment, the spin anisotropy (spheres with arrows in Figs. 2f-h) is governed by the interplay between SOC and Hund's interactions [27], and can therefore vary significantly between the different states. To investigate isospin anisotropy, we image $B_z^{ac}(x, y)$ while rotating $B_a$.

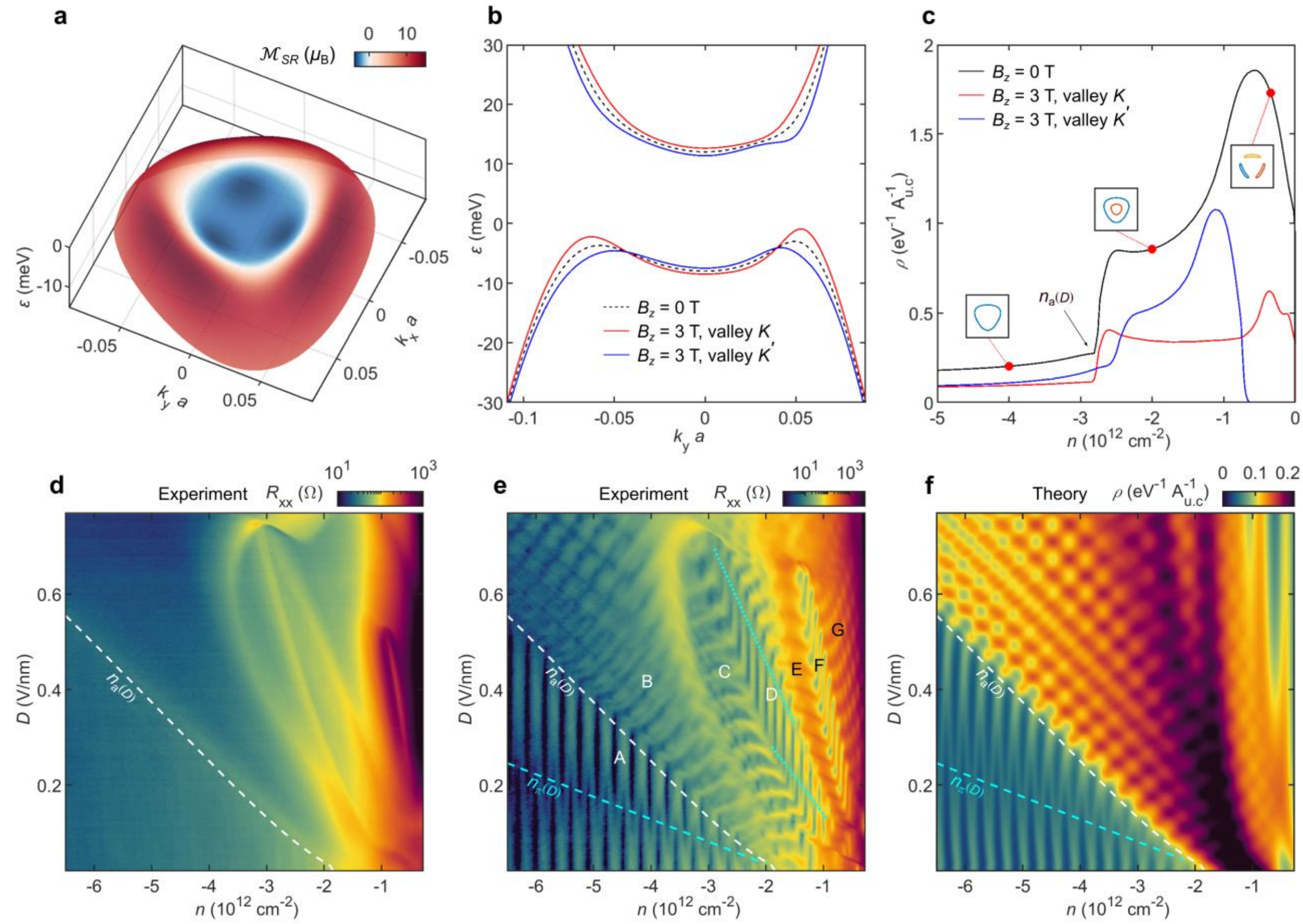

**Fig. 1. Topological magnetic band reconstruction (TMBR) and transport measurements in ABCA graphene. a,** Surface plot of the calculated low-energy band structure of the valence band $\varepsilon(k)$ with overlaid color-coded self-rotation magnetization $\mathcal{M}_{SR}(k)$ of $K$' valley for displacement potential $\Delta_1 = 10$ meV at $B_z = 0$ (Methods). **b,** Energy dispersion, $\varepsilon(k) = \varepsilon_0(k) - \mathcal{M}_{SR}(k)B_z$, along $(0, k_y)$ for $\Delta_1 = 10$ meV at $B_z = 0$ T (dashed) and $B_z = 3$ T, resulting in TMBR and degeneracy lifting between $K$ (red) and $K$' (blue) valleys. **c,** Corresponding valence band DOS, $\rho$, vs. $n$ without (black) and with valley degeneracy lifting by TMBR for $K$ (red) and $K$' (blue) valleys. A full valley polarization occurs at $|n| < 0.76{\times}10^{12}$ $cm^{-2}$ at $B_z = 3$ T. The insets show the Fermi surface structure at the red points and the black arrow indicates the sharp step in DOS at annulus opening $n_a(D)$. **d,** Longitudinal resistance $R_{xx}(n, D)$ of device A at $B_z = 0$ T at $T = 500$ mK. White dashed line is the theoretical fit of the Lifshitz transition $n_a(D)$. **e,** Same as (d) at $B_z = 3$ T showing different patterns of QOs in the various labeled regions. Region A: full metal with single Fermi surface s-M, B: a-M with annular Fermi surface and partial TMBR-induced valley polarization, C: full metal with spontaneous partial spin polarization and TMBR-induced valley polarization, D: valley-degenerate spin-polarized half metal with single Fermi surface s-1/2M, E: a-1/2M with annular Fermi surface and possible partial valley polarization, F: quarter metal 1/4M with single Fermi surface, G: 1/4M with annular Fermi surface. The cyan dashed curve in region A is the $\pi$-shift $n_\pi(D)$ line in full metal derived from the TMBR calculations with no fitting parameters. The cyan dotted curves in region D mark equivalent $\pi$-shift lines in the 1/2M phase. **f,** Calculated DOS showing oscillations due to LLs at $B_z = 3$ T in presence of TMBR. The single particle DOS well reproduces the QOs in regions A and B in (e) where interaction effects are negligible ($D = 1$ V/nm corresponds to $\Delta_1 = 97$ meV).

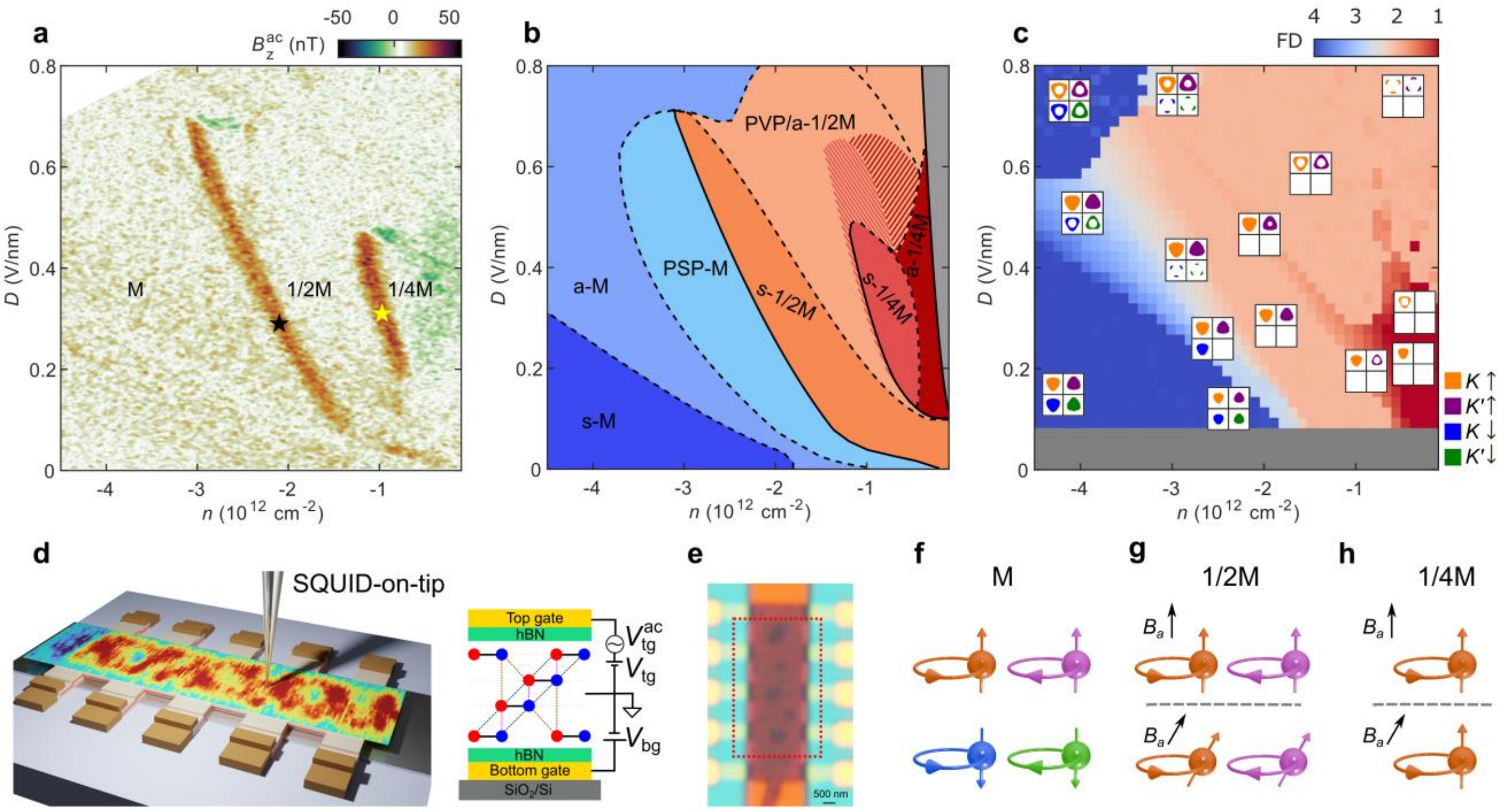

**Fig. 2. Symmetry-broken isospin phase diagram in ABCA graphene. a,** $B_z^{ac}(n, D)$ measured in the bulk of the sample at $T = 20$ mK and $B_z = 10$ mT, showing sharp differential magnetic signal along M-to-1/2M and 1/2M-to-1/4M phase transition lines. **b,** Schematic experimentally derived phase diagram with labeled states. s-M: four-fold degenerate full metal with simple Fermi surface; a-M: full metal with annular Fermi surface; PSP-M: partially spin polarized M with TMBR-induced valley polarization; s-1/2M: two-fold degenerate fully spin polarized 1/2M with simple Fermi surface; PVP-1/2M: partially valley polarized annular 1/2M; s-1/4M: one-fold degenerate simple 1/4M; a-1/4M: annular 1/4M. Solid lines indicate sharp magnetic transitions and dashed lines represent continuous transitions. The striped red-orange regions show schematically the expansion of the s-1/4M and a-1/4M phases upon increasing $B_z$ (Extended Data Figs. 5 and 6). **c**, Calculated scHF phase diagram with color bar indicating flavor degeneracy, $FD$. The squares show the corresponding derived ground state fermiology with colored areas representing hole-occupied states of the four flavors. In the gray area at low $D$ the numerical calculations are unreliable due to overlap of conduction and valence bands. **d**, Schematic sample layout with scanning SOT revealing the local magnetization pattern (left). Schematic cross section of the encapsulated ABCA graphene with electrical diagram (right). **e**, Optical image of the ABCA graphene device B patterned into Hall bar geometry with scan window indicated by the red dotted rectangle. The dark circular patches in the central region are bubbles. **f-h**, Isospin schematics with orbital (circular trajectories) and spin (balls with arrows) sectors. **f**, Four-fold degenerate M with zero magnetization. **g**, Two-fold valley-degenerate spin-polarized 1/2M in out-of-plane (top) and tilted $B_a$ (bottom), with isotropic magnetization and spin direction following $B_a$ orientation. **h**, One-fold spin- and valley-polarized 1/4M in out-of-plane (top) and tilted $B_a$ (bottom). The isospin magnetization is highly anisotropic with easy $\hat{z}$ axis and spin locked to orbital moment.

### Isospin anisotropy in half metal

Figure 3g shows $B_z^{ac}(x, y)$ measured in $B_z = 15$ mT ($B_x = 0$ mT, $\tan\theta_B = B_x/B_z = 0$) across the M-to-1/2M phase transition. Using an inversion procedure (Methods), we reconstruct the local differential magnetization $m(x, y)$ shown in Fig. 3j. Since $n_{ac}$ modulation induces *ac* switching between adjacent phases, the differential magnetization, $\boldsymbol{m} = \boldsymbol{\mathcal{M}}_{1/2M} - \boldsymbol{\mathcal{M}}_M$, reflects the difference in the local magnetization between the 1/2M and M states (for clarity, the differential $m$ is normalized by $n$ in units of $\mu_B$/e). Since M state has no magnetization, the measured $m$ reflects the magnetization of the 1/2M.

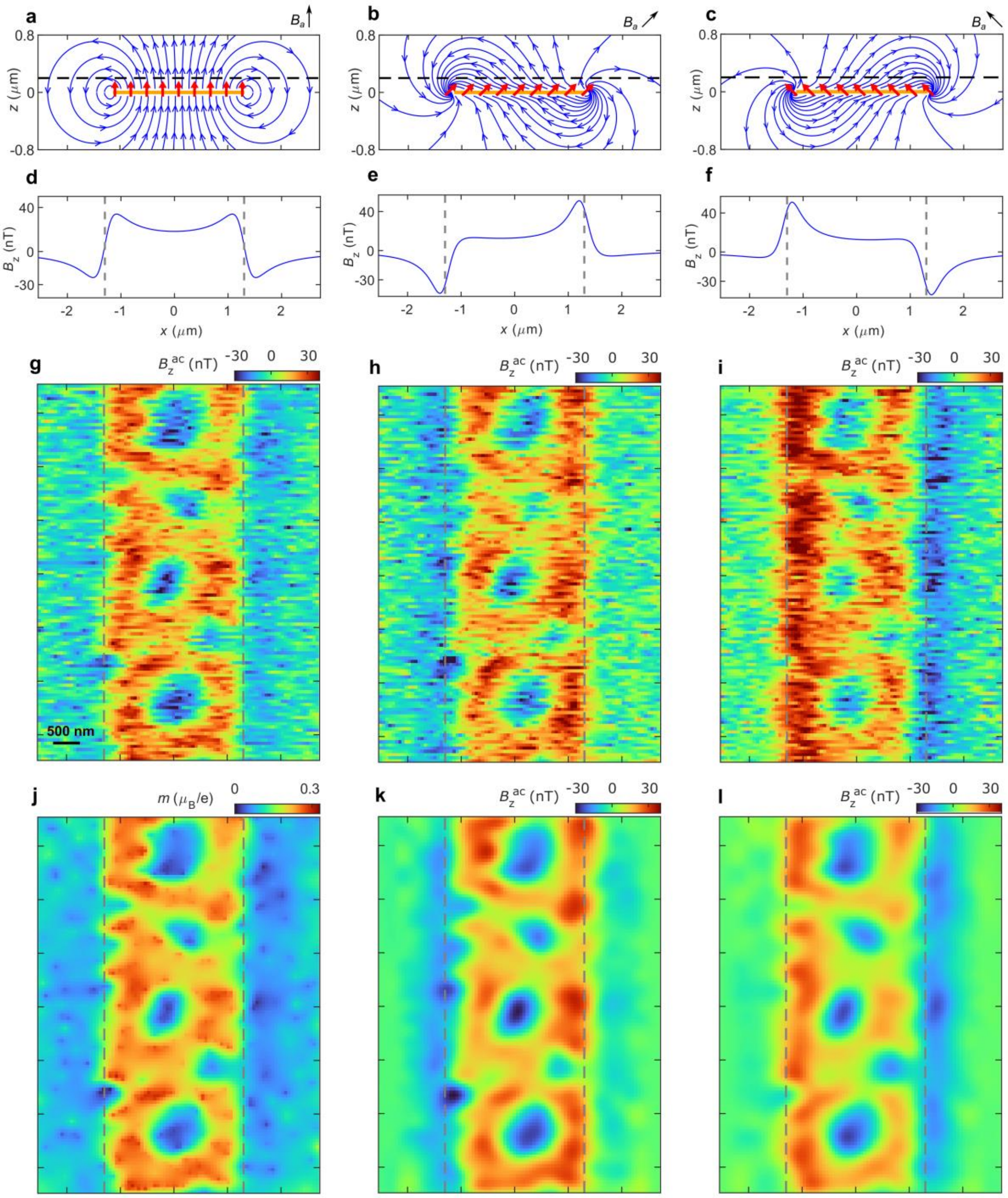

**Fig. 3. Isospin texture in the 1/2M phase. a,** Schematic cross section of graphene (orange) with isospin magnetic moments (red) oriented along $\hat{z}$ ($\theta_s = 0$) with the corresponding calculated stray magnetic field lines. **b,c,** Same as (a) for isospins tilted at $\theta_s = 45°$ (**b**) and $\theta_s = -45°$ (**c**). **d,** Calculated out-of-plane stray magnetic field $B_z(x)$ at a height of 200 nm above the graphene (dashed line in (a)) for uniform magnetization of 0.63×10$^{12}$ $\mu_B$cm$^{-2}$. The vertical dotted lines correspond to graphene edges. **e,** Same as (d) for $\theta_s = 45°$. The positive $B_z$ peak resides at the inner side of the right graphene edge, while the negative peak resides at the outer side of the left edge. **f,** Same as (e) for $\theta_s = -45°$. **g,** Measured out-of-plane stray field $B_z^{ac}(x, y)$ at the M-to-1/2M transition (black star in Fig. 2a) for $B_a$ at $\theta_B = 0$. The blue circular regions correspond to the locations of the bubbles (Fig. 2e). **h,** Same as (g) for $\theta_B = 45°$. The negative $B_z^{ac}$ is enhanced outside the left edge, while the positive $B_z^{ac}$ is enhanced on the inner side of the right edge. **i,** Same as (h) for $\theta_B = -45°$. **j,** Local differential magnetization $m(x, y)$ reconstructed from (g). **k,** Calculated $B_z^{ac}(x, y)$ from $m(x, y)$ in (j) tilted to $\theta_s = 45°$, reproducing the measured $B_z^{ac}(x, y)$ in (h). **l,** Same as (k) for $\theta_s = -45°$.

Next, applying $B_x = 15$ mT rotates the field to $\theta_B = 45°$, with the resulting $B_z^{ac}(x, y)$ shown in Fig. 3h. Although the overall $B_z^{ac}(x, y)$ looks similar, the stray field along the inner right edge is enhanced (red) while the outer edge field vanishes (green), rather than being negative (blue) as in Fig. 3g. In contrast, on the left edge, the outer field becomes more negative and the inner field is reduced. Additionally, bubble domains with suppressed $B_z^{ac}$ become more oval and shift rightward. For $\theta_B = -45°$, similar effects but with reversed orientation appear (Fig. 3i). These changes are consistent with calculated stray fields for uniform sample with tilted magnetization (Figs. 3b,c,e,f).

The data show that the isospin orientation in the 1/2M state depends on the direction of $B_a$ (Fig. 2g). To quantify the magnetization angle, we begin with the reconstructed $\boldsymbol{m}(x, y) = m(x, y)\hat{z}$ at $\theta_B = 0$ (Fig. 3j), rotate int by an angle $\theta_s$, such that $\boldsymbol{m}(x, y) = m(x, y)(\cos\theta_s\,\hat{z} + \sin\theta_s\,\hat{x})$, and compute the resulting $B_z^{ac}(x, y)$. The actual isospin magnetization angle $\theta_s$ is then extracted by best fitting the calculated and measured $B_z^{ac}(x, y)$ (Extended Data Fig. 8). The calculated $B_z^{ac}(x, y)$ (Fig. 3k) well reproduces the measured one (Fig. 3h), including the negative $B_z^{ac}$ along the left edge, the enhanced field at the right edge, and the bubble distortions. A similarly accurate fit for $\theta_B = -45°$ is shown in Fig. 3l. This approach enables extraction of both the magnitude and orientation of the isospin magnetic moment (Methods).

Figure 4f shows $\theta_s$ vs. $\theta_B$ revealing that in the 1/2M state, isospin magnetism is nearly isotropic and aligns closely with $B_a$. Given the strong $\hat{z}$-axis anisotropy of orbital magnetism, this confirms that 1/2M in tetralayer rhombohedral graphene is a valley-degenerate, spin-polarized ferromagnet, consistent with prior RMG studies [15,30,31]. However, full spin isotropy in the presence of SOC implies a large inter-valley Hund's coupling, as discussed below.

For $B_a$ along $\hat{z}$ ($\theta_B = 0$), the symmetric negative stray field $B_z^{ac}$ outside the right and left edges (Fig. 3g) is symmetric, matches calculations for $m(x, y)$ aligned along $\hat{z}$ (Figs. 3a,d; Methods). The reconstructed $m(x, y)$ is rather uniform in the sample bulk (red-yellow in Figs. 3g,j), except for circular patches (blue) where the magnetization vanishes. These are the bubbles formed during fabrication (Fig. 2e) that induce strain and disorder that destroy the symmetry-broken phases, as seen in magic-angle bilayer graphene [50].

**Anisotropy in quarter metal**

Figure 2a reveals a sharp peak in differential magnetization at the 1/2M-to-1/4M transition, indicating significantly greater magnetization in the 1/4M phase. Since the 1/2M is already spin polarized, this increase must stem from additional valley polarization. To probe the isospin structure, we image $B_z^{ac}(x, y)$ across the transition. Figure 4a shows the data at $\theta_B = 0°$, with the reconstructed $m(x, y)$ in Fig. 4d. The $m(x, y)$ pattern resembles that of the M-to-1/2M transition but with over threefold higher amplitude, exceeding 1 $\mu_B$/e. Since for $\theta_B = 0°$ the spin is aligned along $\hat{z}$ with the same contribution to magnetization both in the 1/2M and 1/4M states, the observed differential $m(x, y)$ across the transition reflects only the Berry-curvature-induced orbital magnetization in the 1/4M, emphasizing its dominant contribution.

Rotating $B_a$ to $\theta_B = 83°$, the resulting $B_z^{ac}(x, y)$ in Fig. 4b shows the negative stray field shifting to the right edge—opposite to its position in Figs. 3e,h—suggesting that the differential magnetization behaves as if tilted to negative $\theta_s$ (Figs. 3f,i). This counterintuitive behavior reveals a strong SOC in the 1/4M phase, aligning both spin and orbital moments along $\hat{z}$. As a result, the difference between the out-of-plane magnetization in 1/4M and the $\theta_s = 83°$ spin tilt in the 1/2M results in differential magnetization that appears to a negative $\theta_s$. Similarly, for $\theta_B = -83°$ the differential magnetization appears to be oriented in the positive direction with negative $B_z^{ac}$ on the left edge (Fig. 4c). As discussed in Methods, we can qualitatively reproduce the measured $B_z^{ac}(x, y)$ in Fig. 4b by proper superposition of orbital and spin tilted $m(x, y)$ patterns from Figs. 4d and 3j, as shown in Fig. 4e. These observations show that, unlike the nearly isotropic 1/2M, the 1/4M phase exhibits strong isospin anisotropy with easy $\hat{z}$ axis, as shown by the magenta triangles in Fig. 4f (Extended Data Fig. 9).

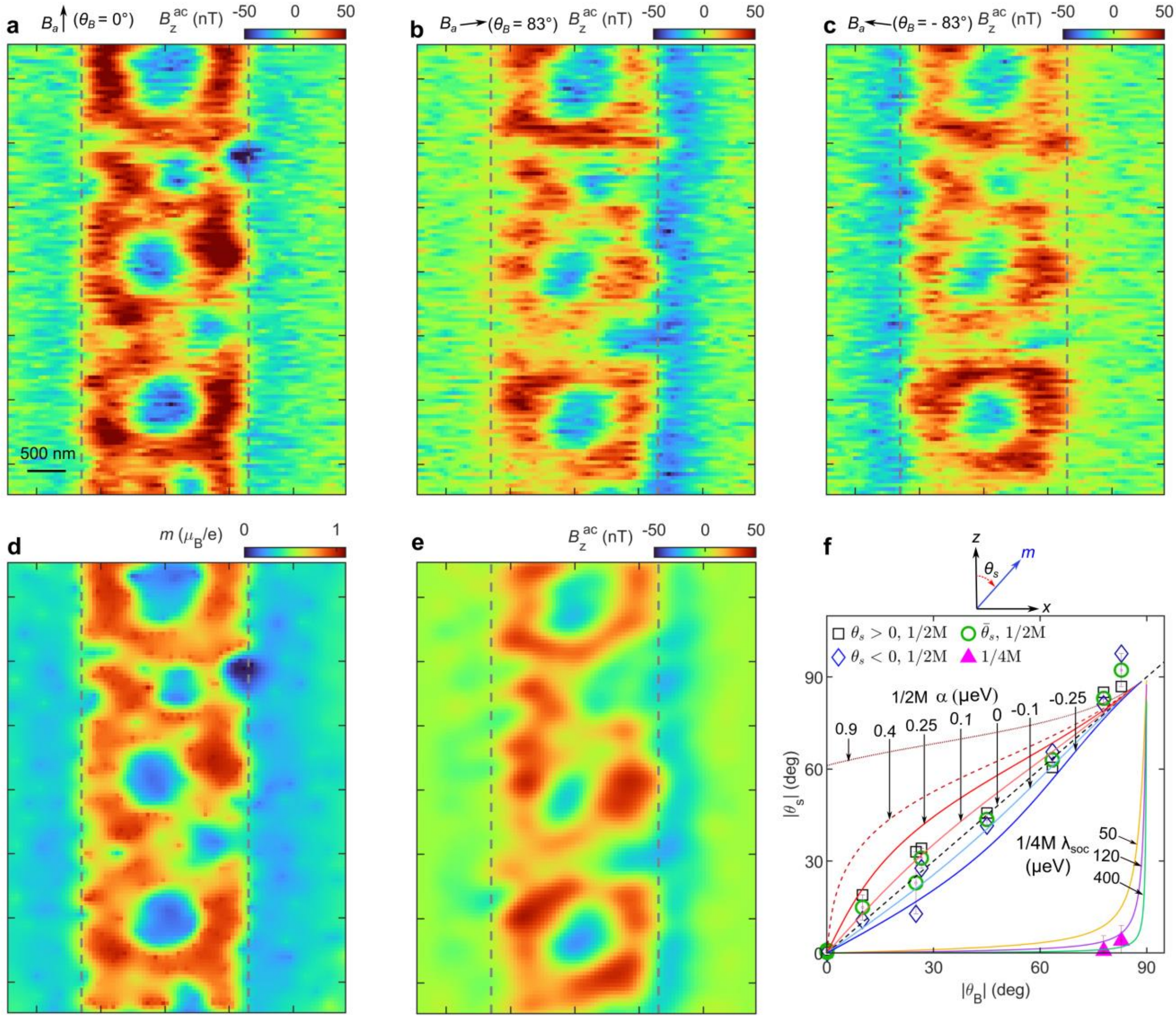

**Fig. 4. Isospin texture in the 1/4M phase. a,** Measured $B_z^{ac}(x,y)$ at the 1/2M-to-1/4M transition (yellow star in Fig. 2a) for $B_a$ along $\theta_B = 0$. **b,** Same as (a) for $\theta_B = 83°$. In contrast to Fig. 3h, the negative $B_z^{ac}$ occurs at the right graphene edge due to spin tilting in the 1/2M phase. **c,** Same as (b) for $\theta_B = -83°$. **d,** Local differential magnetization $m(x,y)$ reconstructed from (a). **e,** $B_z^{ac}(x,y)$ calculated using a superposition of the orbital $m(x,y)$ in the 1/4M shown in (d) and of a tilted spin $m(x,y)$ in the 1/2M based on Fig. 3j, qualitatively reproducing the measured $B_z^{ac}(x,y)$ in (b). **f,** The derived spin tilt angle $|\theta_s|$ vs. the applied field tilt angle $|\theta_B|$ for positive (black squares) and negative (blue diamonds) angles across the M-to-1/2M and 1/2M-to-1/4M (solid magenta triangles) transitions. The green open circles represent the average spin tilt angle, $\bar{\theta}_s = (|\theta_s(>0)| + |\theta_s(<0)|)/2$. Black and magenta line bars represent corresponding error in the spin tilt angle. The solid lines show the calculated $\theta_s$ vs. $\theta_B$ (Methods) for various indicated values of $\lambda_{SOC}$ in 1/4M and of $\alpha$ in 1/2M for $B_z = 15$ mT and variable $B_x = B_z \tan\theta_B$. The 1/4M lines set a lower bound on Ising SOC of $\lambda_{SOC} \gtrsim 120$ μeV substantially higher than the previously evaluated $\lambda_{SOC} = 50$ μeV (yellow curve) in ABC graphene [27]. The magnetism in the 1/2M is essentially fully isotropic with the calculated solid lines setting an upper bound on spin anisotropy of $|\alpha| < 0.25$ μeV, which sets a lower bound on $U_{Hu} \gtrsim 6.5$ meV (Methods). The dashed red line shows calculated $\theta_s$ vs. $\theta_B$ for $\alpha = 0.4$ μeV based on previously assumed $U_{Hu} = 4.2$ meV [42,51], which is inconsistent with the data.

**Analysis of isospin anisotropy and Hund's and SOC energies**

The stark anisotropy difference between the 1/2M and 1/4M can be qualitatively understood as follows. In the 1/4M, the spin anisotropy is dictated by SOC energy $\lambda_{SOC}$ which tends to align the spin and valley moments. Considering SOC and Zeeman effects under tilted fields, the spin tilt angle is given by $\tan\theta_s = \frac{B_x}{B_z+\lambda_{SOC}/(g\mu_B)}$ (Methods), with $g$ being the gyromagnetic ratio. The three rightmost curves in Fig. 4f show the calculated $\theta_s$ vs. $\theta_B = \tan^{-1}(B_x/B_z)$ for different values of $\lambda_{SOC}$ at $B_z = 15$ mT and variable $B_x$, which allows us to set a lower bound on Ising $\lambda_{SOC} \geq 120$ µeV. This experimental estimate of the SOC strength in ABCA is more than twice the previously evaluated $\lambda_{SOC} = 50$ µeV in ABC graphene [27].

The extracted $\lambda_{SOC}$ raises a question regarding the isotropic spin moment measured in the 1/2M phase. A finite SOC means that full spin rotation is not a symmetry of the system, and hence we expect a spin ferromagnetic phase to be anisotropic (either easy-axis or easy-plane). However, the spin anisotropy $\alpha$ in the 1/2M is significantly suppressed by the Hund's coupling, $\alpha \cong \lambda_{SOC}^2/8U_{Hu}$, where $U_{Hu} = n \cdot J_H$ is the intervalley Hund's coupling energy (Methods). Therefore, to explain the seemingly isotropic spin moment, $U_{Hu}$ should be sufficiently large. The 1/2M solid lines in Fig. 4f show the calculated $\theta_s$ vs. $\theta_B$ dependence for various values of $\alpha$ (Extended Data Fig. 7), from which we can set an upper bound on $|\alpha| < 0.25$ µeV. Using the attained $\lambda_{SOC} > 120$ µeV, this $\alpha$ leads to a lower bound on the Hund's coupling of $U_{Hu} > 6.5$ meV $= 75.4$ K, and of the corresponding $J_H > 3.1\times10^{-12}$ meV·cm$^2$. This measurement of the Hund's type intervalley exchange energy is significantly larger than the values of $J_H \cong 2\times10^{-12}$ meV·cm$^2$ previously used in literature for modeling rhombohedral graphene [27,42,51]. Such large Hund's energy scale means that relative spin fluctuations between the two valleys are suppressed. For temperatures below this energy scale, the total spin moment for the two valleys should be treated as a single degree of freedom, while at higher temperatures the spin moments associated with each valley should be considered as independent degrees of freedom.

**Phase diagram**

Low-field magnetization data (Fig. 2a) reveal three main phases, M, 1/2M, and 1/4M, separated by sharp phase transitions. Examination of the QOs at high magnetic fields in Fig. 1e and Extended Data Fig. 4, in contrast, discloses additional symmetry broken states within the main phases as summarized in Fig. 2b. The M phase, in which all four flavors are occupied, is subdivided into three states: simple s-M with single Fermi surface, a-M with annular Fermi surface, and spontaneously symmetry broken partially spin polarized PSP-M state. Since PSP state is partially polarized, the sharp transition from PSP-M to fully-spin-polarized 1/2M should be accompanied by a reduced jump in magnetization, in accord with the observed $m \cong 0.3$ $\mu_B$/e at the M-to-1/2M transition. The 1/2M phase is subdivided into valley symmetric s-1/2M with simple Fermi surface and a-1/2M with annular Fermi surface which is possibly also a partially valley polarized PVP-1/2M. Note that similarly to full metal, the TMBR-induced $\pi$ shifts are also observed in the s-1/2M (dotted cyan lines in Fig. 1e), revealing magnetic-field-induced partial valley polarization. Finally, the 1/4M phase is also subdivided into a single Fermi surface s-1/4M and annular a-1/4M states. Since 1/4M comprises of only one valley and one spin, no $\pi$ shifts should occur. Notably, the 1/4M region expands markedly with field (striped regions in Fig. 2b) as seen by comparing Figs. 1d and 1e and described in Extended Data Fig. 6. This expansion is a result of magnetic energy gain due to orbital magnetic moment in the 1/4M, which is absent in the adjacent 1/2M (Methods). These conclusions are corroborated by self-consistent Hartree-Fock (scHF) calculations with gate-screened Coulomb interaction (Methods). Figure 2c shows the derived phase diagram with color-coded flavor degeneracy, $FD = n/\max(n_i)$, where $n_i$ is the carrier density in each isospin flavor, showing a sequence of symmetry-breaking transitions from M to 1/4M, which broadly agrees with the measured diagram in Fig. 2b.

## Discussion

A comparison of isospin transitions in ABCA and ABC graphene [27] shows that in both systems, full spin polarization does not occur immediately. Instead, there is a gradual polarization, too weak to be detected locally, followed by an abrupt transition to a fully spin-polarized 1/2M phase. Similarly, in presence of magnetic field the spin-polarized 1/2M apparently undergoes gradual valley polarization, before an abrupt transition to a 1/4M. Notably, while ABC graphene exhibits IVC ordering within the 1/4M, ABCA shows no such signatures, either in measurements or scHF simulations. This is particularly interesting since hole-doped ABC graphene has been found to become superconducting at sub-Kelvin temperatures [23], while thus far no superconducting phase has been reported in pure hole doped tetralayer [43]. The ABCA phase diagram thus presents a valuable comparison for narrowing down the superconducting mechanism. Importantly, the hole-doped ABCA seems to be lacking IVC phases and its superconductivity is apparently suppressed, while in ABC both phases are present, which is consistent with recent suggestions of a pairing mechanism mediated by IVC fluctuations [10,11,14,25]. Since our findings clearly indicate that the isospin anisotropy throughout the phase diagram depends subtly on small differences in microscopic properties, a better understanding of the underlying electronic correlations is called for.

The findings not only shed light on the microscopic aspects of isospin ordering in fine detail, but also demonstrate that RMG is a highly tunable and increasingly well-understood system with superior device quality compared to twisted stackings. Recent work on spin-orbit proximitized RMG has uncovered new superconducting phases absent without strong SOC [37–42], potentially originating from a spin-canted in-plane ferromagnetic state [26,42]. Our experimental estimate of Hund's coupling is crucial for quantitative understanding of spin-canted ferromagnetism and possibly related superconducting phases. We conclude that screening effects on the Hund's coupling are weak, as the attained $U_{Hu} > 6.5$ meV is suppressed by less than a factor of two relative to the maximal value of intervalley Coulomb energy in vacuum (Methods). Comparable Hund's coupling is thus expected across different rhombohedral graphene systems.

**Acknowledgments** We thank Andrea F. Young for fruitful discussions. This work was co-funded by the Minerva Foundation grant No 140687, by the United States - Israel Binational Science Foundation (BSF) grant No 2022013, by the Institute for Artificial Intelligence, and by the European Union (ERC, MoireMultiProbe - 101089714). Views and opinions expressed are however those of the author(s) only and do not necessarily reflect those of the European Union or the European Research Council. Neither the European Union nor the granting authority can be held responsible for them. E.Z. acknowledges the support of the Andre Deloro Prize for Scientific Research, Goldfield Family Charitable Trust, and Leona M. and Harry B. Helmsley Charitable Trust grant #2112-04911. Y.O. acknowledges support from the European Union's Horizon 2020 research and innovation programme (Grant Agreement LEGOTOP No. 788715), the DFG (CRC/Transregio 183, EI 519/7-1), and the Israel Science Foundation ISF (Grant No 1914/24). E.B. acknowledges support from the European Research Council (ERC) under grant HQMAT (Grant Agreement No. 817799), and from an NSF-BSF award DMR-2000987. T.H. acknowledges financial support by the European Research Council (ERC) under grant QuantumCUSP (Grant Agreement No. 101077020). K.W. and T.T. acknowledge support from the JSPS KAKENHI (Grant Numbers 21H05233 and 23H02052), the CREST (JPMJCR24A5), JST and World Premier International Research Center Initiative (WPI), MEXT, Japan.

**Author contributions** N.A. designed and built the scanning SOT microscope and S.G. advanced the software. S.D. and N.A. performed the local magnetization studies. M.U. fabricated and characterized the samples. M.U., N.A., and S.D. performed the transport measurements and data analysis. S.D., N.A., M.U. and E.Z. designed the experiment. S.D. and Y.M. fabricated the SOT and the tuning fork, and M.E.H. developed the SOT readout. Y.V., T.H., P.E., Y.O. and E.B. performed the theoretical modeling. Y.V., S.D. Y.Z., and N.A. performed the scHF, Stoner, and phase diagram calculations. Y.V. and S.D. performed the Hund's and SOC modeling and calculations. S.D., Y.Z., and P.E. performed the LL calculations. A.Y.M. developed the magnetization reconstruction algorithm and S.D., Y.Z., and A.Y.M. carried out the anisotropy analysis. K.W. and T.T. provided the hBN crystals. S.D., N.A, M.U., Y.V., T.H. and E.Z. wrote the original manuscript. All authors participated in discussions and revisions of the manuscript.

**Competing interests** The authors declare no competing interests.

**Data availability** The data that supports the findings of this study are available from the corresponding authors on reasonable request.

**Code availability** The scHF, anisotropy, and magnetic inversion calculations used in this study are available from the corresponding authors on reasonable request.

## Methods

### Device fabrication

The hBN-encapsulated tetralayer graphene heterostructures were fabricated using the dry-transfer method. The hBN and graphene flakes were exfoliated onto a Si/SiO$_2$ (285 nm) substrate and picked up using a polycarbonate (PC) on a polydimethylsiloxane (PDMS) dome stamp. The number of layers of the graphene flakes was determined via optical contrast analysis against the Si substrate (using the G values in RGB triplet) and further confirmed through step strength analysis in dark field mode. The stacking order of the graphene flakes was determined by the shape of the Raman 2D peak (Extended Data Fig. 1e), and its rhombohedral-stacked regions were isolated using high-power laser cutting. The Raman analysis was subsequently repeated on the final stack (Extended Data Fig. 1f) to rule out possible relaxation of the graphene to Bernal stacking.

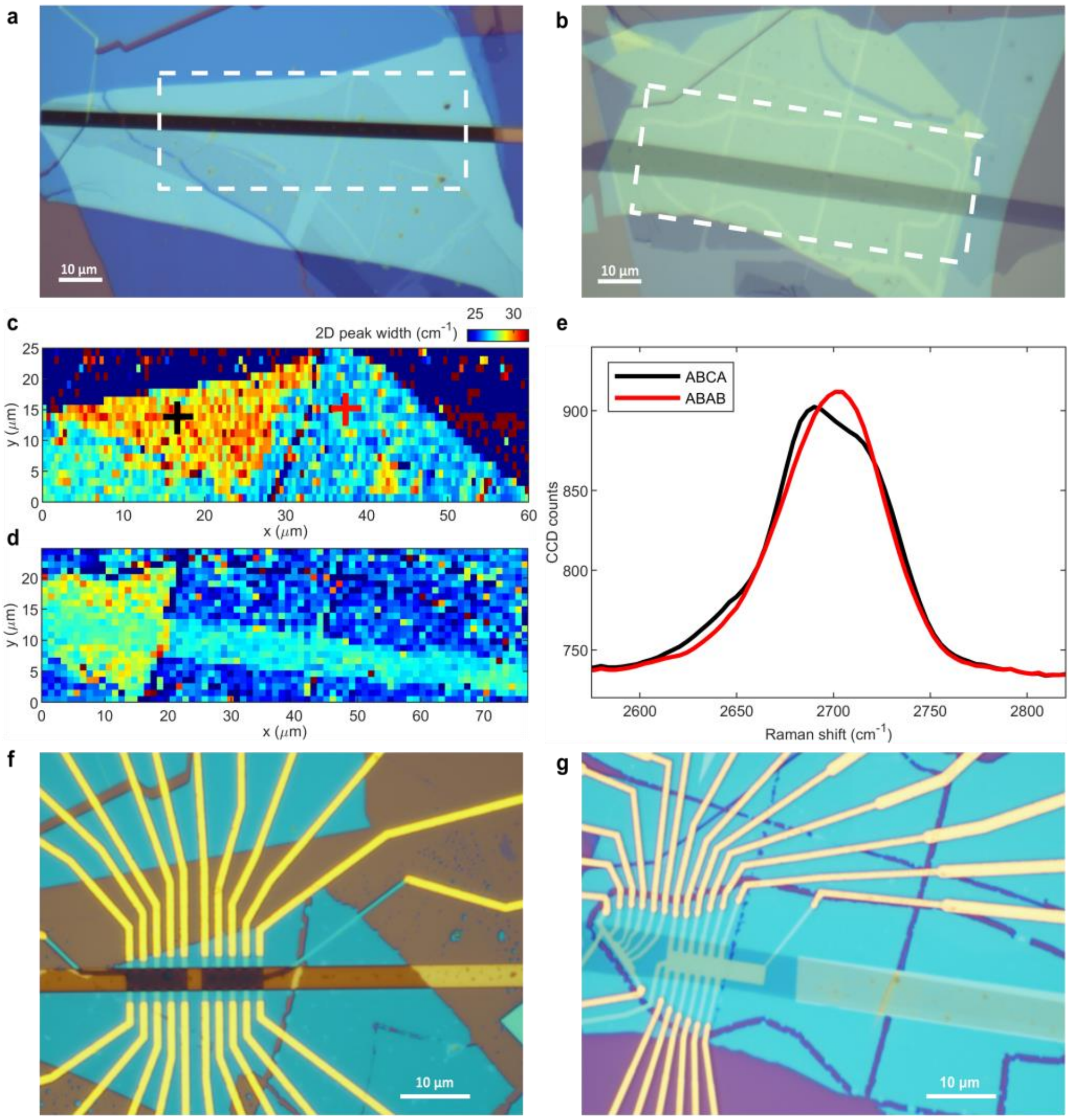

**Extended Data Fig. 1. Device fabrication of rhombohedral graphene. a,** Optical image of the hBN/ABCA Graphene/hBN stack (blue) on Si/SiO$_2$ substrate (gray-brown) with Ti/Pt bottom gate (dark brown) for device B. **b,** Same as (a) for devices A and C. **c,** Raman map of the Gaussian-fitted width of the 2D peak in the dashed rectangle in (a). The yellow-red triangular region corresponds to ABCA graphene stacking, and the blue-green regions correspond to ABAB stacking, separated by sharp domain walls. **d,** Same as (c) for the region marked by the dashed rectangle in (b). **e,** Typical 2D peaks of the Raman spectrum for the two stacking configurations measured at the points marked by black and red crosses in (c). The ABAB 2D peak is narrower and more symmetric than the ABCA 2D peak. **f,** Optical image of the final device B (right Hall bar). **g,** Final devices A (left) and C (right).

To increase the yield of the rhombohedral structure, a single-pick strategy was used to avoid multiple pick-ups of graphene flakes during the stacking procedure which enhance relaxation to Bernal stacking. This involved pre-preparation of the lower part of the stack, comprising of a backgate (annealed patterned metal or graphite stripe) covered with an hBN flake, followed by annealing at 500 °C under vacuum. For the same reason, the graphene flake manipulations were done faster, at lower temperatures (<70 °C) than usual, and along the zigzag direction of the graphene [52]. This modified stacking protocol can, in turn, encourage bubble formation, as can be seen in Fig. 2e. Unlike in other graphene stacks, the bubbles cannot be removed by thermal annealing or subsequent pick-ups since the graphene stacking configuration is susceptible to these procedures. The chirality of the edges (zigzag/armchair) was first determined from the dependence of the second harmonic generation (SHG) on the polarization angle in the hBN flakes [53,54] and in Bernal-stacked trilayer region within the same flake as the tetralayer graphene of interest.

Subsequently, the top hBN and graphene flakes were transferred onto the pre-prepared bottom part. During these processes, the crystal axes (discerned from straight edges) of the hBN and tetralayer graphene were intentionally misaligned using a mechanical rotation stage to avoid moiré correlations. A Ti (2 nm)/Au (10 nm) top gate was deposited onto the finalized stacks. Then, the devices were etched into a Hall bar geometry, and 1D contacts were established through $SF_6$ and $O_2$ plasma etching, followed by Cr (4 nm)/Au (60-90 nm) evaporation. The Raman, SHG measurements and the laser cutting were performed using a WITec alpha300 R Raman imaging microscope using 532 and 1064 nm wavelengths.

Device summary:

**Devices A and C**: Ti (2 nm) / Au (10 nm) top gate and graphite bottom gate, the top and bottom hBN thicknesses are $\cong$ 17 nm and 31 nm respectively.

**Device B**: Ti (2 nm) / Au (10 nm) top gate and Ti (2 nm) / Pt (10 nm) bottom gate, the top and bottom hBN thicknesses are $\cong$ 18 nm and 29 nm respectively.

**SOT measurements**

The local magnetic imaging was performed using a home-built scanning SOT microscope in a dilution refrigerator. All measurements were conducted at the nominal base temperature of $T = 20$ mK. For detecting the local magnetic fields, we employed indium SQUID fabricated on the tip of a pulled quartz pipette with an integrated shunt resistor [55]. The effective diameter of the In SOT was 194 nm with field sensitivity down to 10 nT/$Hz^{1/2}$, and operating fields up to 0.4 T. A cryogenic series SQUID array amplifier was implemented to read out the SOT signal [56,57]. To control the scan height during imaging, the SOT was attached to a quartz tuning fork vibrating at a resonance frequency of around 32 kHz. A small *ac* voltage $V_{tg}^{ac}$ was applied to the top gate with amplitude of 100 to 200 mV (rms) and frequency of 0.7 to 2 kHz, modulating the carrier density by $n_{ac}$, and the corresponding *ac* change in the local magnetic field $B_z^{ac}(x, y)$ was acquired by the SOT.

All the presented $B_z^{ac}(x, y)$ images were measured on device B to avoid magnetic signals originating from the graphite bottom gate. Single point measurements (like Fig. 2b) were recorded at a height of 70 nm above the surface of the top metal gate. The 2D spatial images (Figs. 3 and 4) were scanned at a constant height of about 150 nm above the top gate surface, with pixel size of 50 nm and integration time of 1.2 s/pixel. Figures 3g-i, 4a, and 4b,c were acquired at $(D, n)$ values of (0.29 V/nm, $-2.1\times10^{12}$ $cm^{-2}$), (0.31 V/nm, $0.97\times10^{12}$ $cm^{-2}$), (0.31 V/nm, $0.9\times10^{12}$ $cm^{-2}$), respectively.

The angular dependent measurements were performed at fixed $B_z = 15$ mT or 85 mT and variable $B_x$. All the data points in Fig. 4f were acquired at $B_z = 15$ mT except for $\theta_B < 30°$ points, which were measured also at $B_z = 85$ mT.

### Single particle band structure calculations

To calculate the low energy single particle band structure of rhombohedral tetralayer graphene, we adopted Slonczewski-Weiss-McClure (SWMC) parameterized tight-binding model as given in [12]. Using the sublattice states {A1, B1, A2, B2, A3, B3, A4, B4}, the low energy Hamiltonian can be written as,

$$H_0 = \begin{bmatrix} \Delta_1+\Delta_2 & v_0\pi^\dagger & v_4\pi^\dagger & v_3\pi & 0 & \frac{1}{2}\gamma_2 & 0 & 0 \\ v_0\pi & \Delta_1+\Delta_2+\delta & \gamma_1 & v_4\pi^\dagger & 0 & 0 & 0 & 0 \\ v_4\pi & \gamma_1 & \frac{\Delta_1}{3}-\Delta_2+\delta & v_0\pi^\dagger & v_4\pi^\dagger & v_3\pi & 0 & \frac{1}{2}\gamma_2 \\ v_3\pi^\dagger & v_4\pi & v_0\pi & \frac{\Delta_1}{3}-\Delta_2+\delta & \gamma_1 & v_4\pi^\dagger & 0 & 0 \\ 0 & 0 & v_4\pi & \gamma_1 & -\frac{\Delta_1}{3}-\Delta_2+\delta & v_0\pi^\dagger & v_4\pi^\dagger & v_3\pi \\ \frac{1}{2}\gamma_2 & 0 & v_3\pi^\dagger & v_4\pi & v_0\pi & -\frac{\Delta_1}{3}-\Delta_2+\delta & \gamma_1 & v_4\pi^\dagger \\ 0 & 0 & 0 & 0 & v_4\pi & \gamma_1 & -\Delta_1+\Delta_2+\delta & v_0\pi^\dagger \\ 0 & 0 & \frac{1}{2}\gamma_2 & 0 & v_3\pi^\dagger & v_4\pi & v_0\pi & -\Delta_1+\Delta_2 \end{bmatrix}$$

where $\pi = \tau k_x + ik_y$, $v_i = \frac{\sqrt{3}}{2\hbar}a_0\gamma_i$ ($i = 0,3,4$) is the hopping velocity, $\tau$ is the valley index ($\pm1$ for $K$ and $K'$ valleys respectively), and $a_0 = 2.46$ Å is the graphene lattice constant. The parameter $\delta$ corresponds to the on-site potential for the sublattices which have nearest neighbor on adjacent layer. $\Delta_1$ determines the potential difference between the outer layers and is approximately proportional to the applied displacement field $D$, and $\Delta_2$ encodes the potential difference between the outer and the middle layers. The hopping parameters were chosen by fitting the annulus opening line $n_a(D)$ in the full metal in Fig. 1d, which can be achieved with the same intra- and interlayer hopping values reported previously for rhombohedral trilayer graphene [15], leading us to adjust only the value of the on-site energy $\Delta_2$ compared to the trilayer. This results in the following values:

| Parameter | $\gamma_0$ | $\gamma_1$ | $\gamma_2$ | $\gamma_3$ | $\gamma_4$ | $\delta$ | $\Delta_2$ |
|---|---|---|---|---|---|---|---|
| Value (in eV) | 3.1 | 0.38 | -0.015 | -0.29 | -0.141 | 0.0105 | 0.002 |

From the fit to the $\pi$-shift line (Fig. 1), we find that the outer layers potential difference $\Delta_1$= 97 meV corresponds to the applied displacement field $D = 1$ V/nm.

### Topological magnetic band reconstruction

In conventional ferromagnets, magnetic field causes the spin polarized bands to shift in opposite directions in energy due to the Zeeman effect, $\varepsilon = -\boldsymbol{\mathcal{M}} \cdot \boldsymbol{B}$. The topological nature of the graphene bands gives rise to a similar effect, $\varepsilon = -\boldsymbol{\mathcal{M}_{SR}} \cdot \boldsymbol{B}$, where $\boldsymbol{\mathcal{M}}_{SR} = \mathcal{M}_{SR}\hat{z}$ is the self-rotation component of the Berry-curvature-induced orbital magnetization [3]. In contrast to gapped monolayer graphene in which the maxima of Berry curvature $|\Omega(k)|$ and hence of $|\mathcal{M}_{SR}(k)|$ reside at high symmetry points, in rhombohedral graphene at finite $D$, the distribution of $\Omega(k)$ in the momentum space $k$ has doughnut shape (Extended Data Fig. 2), which gives rise to a magnetization with the opposite sign of $\mathcal{M}_{SR}(k)$ in the outer and inner regions (Fig. 1a). As a result, rather than shifting the bands, the magnetic field distorts the band structure.

Using the semiclassical perturbation approach, the modified band energy is expressed as [47]

$$\varepsilon_{n,B}(k) = \varepsilon_{n,0}(k) - \mathcal{M}_{SR,n}(k)B_z,$$

where $\varepsilon_{n,0}$ denotes the n[th] band energy in zero magnetic field. $\mathcal{M}_{SR,n}(k)$ is the orbital magnetization arising from the self-rotation of the Bloch wave function in n[th] band induced by the Berry curvature and is given by [3,8],

$$\mathcal{M}_{SR,n}(k) = -\frac{e}{\hbar} Im\left(\sum_{m\neq n} \frac{< n|\partial_{k_x} H_0|m >< m|\partial_{k_y} H_0|n >}{\varepsilon_{n,0}(k) - \varepsilon_{m,0}(k)}\right)\hat{z},$$

as shown in Fig. 1b. Extended Data Fig. 2a depicts the calculated momentum-space-dependent Berry curvature [3] $\Omega(k)$ and $\mathcal{M}_{SR,n}(k)$ for $\Delta_1 = 10$ meV. Note that in the $K$ valence band, $\mathcal{M}_{SR,v}(k)$ is positive in the vicinity of $k = 0$ (Extended Data Fig. 2b), giving rise to a decrease in $\varepsilon_{n,B}(k)$ at low $k$ in Fig. 1b (red curve), while at larger momenta, the orbital magnetization is negative, thus increasing the band energy and giving rise to nontrivial TMBR. In contrast, in the conduction band, $\mathcal{M}_{SR,c}(k)$ is negative at any $k$ (Extended Data Fig. 2c), resulting in a more uniform upward shift of the $K$ valley conduction band in Fig. 1b. In $K$' valley, the orbital magnetization is of opposite sign, giving rise to pronounced field-induced valley polarization as described in Fig. 1.

Since $K$ and $K$' valleys have opposite $\mathcal{M}_{SR}$, this topological magnetic band reconstruction (TMBR) leads to momentum-dependent valley symmetry breaking and to a striking change in the DOS at elevated fields (red and blue curves in Fig. 1c). The TMBR is strongly pronounced for the annular Fermi surface, causing the LLs of opposite valleys to originate at different densities and interfere with each other, giving rise to a beating pattern (Extended Data Fig. 3). The area of the annular Fermi surface increases with $D$, resulting in more pronounced beating higher displacement fields, as observed in extended Data Figs. 3 and 4. At high carrier densities in the simple Fermi surface region, the TMBR has a more subtle effect shifting the phase of the QOs. When $K$ and $K$' LLs acquire a relative shift of half a period, the amplitude of the QOs is suppressed and a $\pi$-shift in the $n - D$ oscillation pattern occurs as shown in Fig. 1f. The theoretically derived $n_\pi(D)$ line accurately matches the experimental one in Fig. 1e with no additional fitting parameters. Note that the TMBR and the valley degeneracy lifting is a purely single particle effect, which coexists with interaction-driven spontaneous symmetry breaking mechanisms at lower $|n|$.

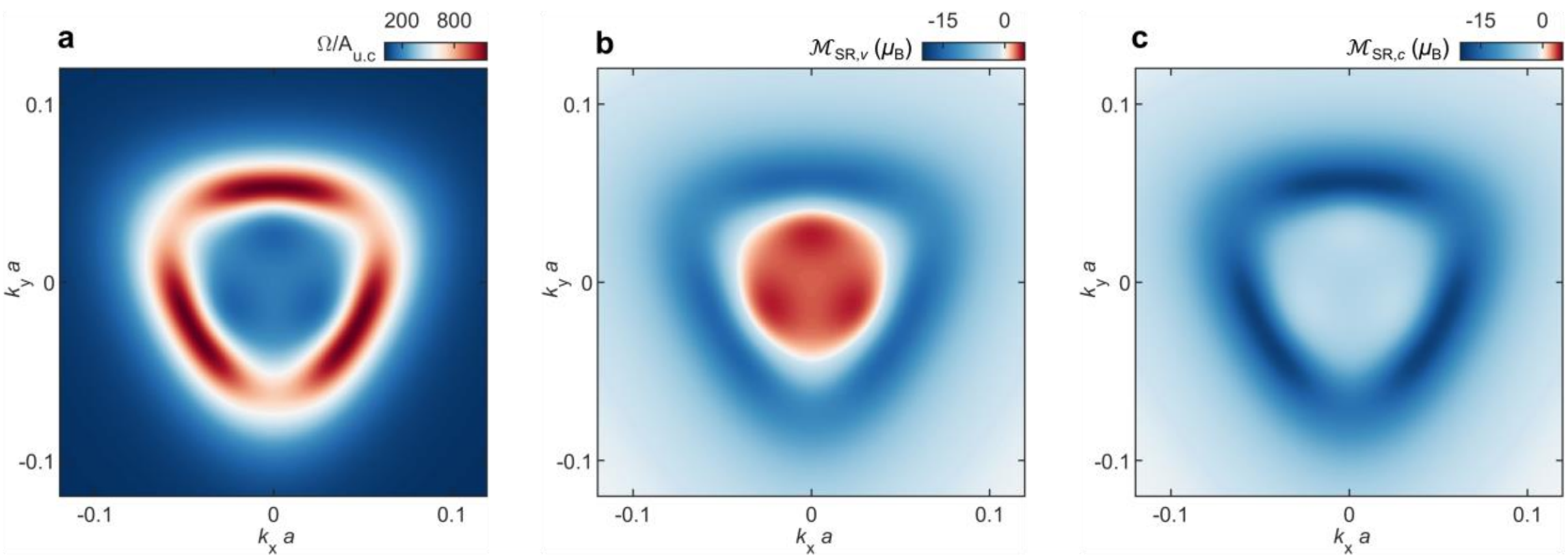

**Extended Data Fig. 2. Berry curvature and orbital magnetism in rhombohedral ABCA graphene. a,** Momentum dependent Berry curvature $\Omega(k)$ in valley $K$ in the valence band for $\Delta_1 = 10$ meV. **b,** Self-rotation component of the orbital magnetization in $K$ valence band, $\mathcal{M}_{SR,v}(k)$, with positive values near zero $k$ and negative values at larger momenta. **c,** Same as (b) in the conduction band, $\mathcal{M}_{SR,c}(k)$, with negative values for all $k$. In $K$' valley the $\Omega(k)$ and $\mathcal{M}_{SR,n}(k)$ are rotated by 180° and have an opposite sign.

Note that TMBR is governed only by the self-rotation term of the Berry-curvature-induced orbital magnetization $\mathcal{M}_{SR}$, while the Chern magnetization term $\mathcal{M}_C$ has no contribution [3,8] since the corresponding energy $-\mathcal{M}_C(k)B_z$ contributes to transfer of electronic states between the topological bands according to Streda's formula, rather than Zeeman-like shifting of the band energy. Note also that the semiclassical approximation calculation is presented here for gaining intuition about the nontrivial effects of the orbital magnetization in a magnetic field. However, the full description in presence of magnetic field should be based on calculation of LLs, as presented below.

**Landau level calculations**

To understand the spectrum of LLs in a perpendicular magnetic field, we consider the theoretical method described in [20]. By implementing the Peierls substitution, the hopping momenta $\pi$ and $\pi^\dagger$ act as rising and lowering operators. In the magnetic oscillator basis $\emptyset_n$, the matrix elements of $\pi$ and $\pi^\dagger$ can be computed for two valleys using the following relations,

$$K: \pi\emptyset_n = \frac{i\hbar}{l_B}\sqrt{2(n+1)}\emptyset_{n+1}$$

$$\pi^\dagger\emptyset_n = -\frac{i\hbar}{l_B}\sqrt{2n}\emptyset_{n-1}$$

$$K': \pi\emptyset_n = \frac{i\hbar}{l_B}\sqrt{2n}\emptyset_{n-1}$$

$$\pi^\dagger\emptyset_n = -\frac{i\hbar}{l_B}\sqrt{2(n+1)}\emptyset_{n+1}$$

where $l_B = \sqrt{\hbar/eB}$ is the magnetic length. To determine the LL energies, $E_i$, we diagonalized $H_0$ by expanding the sublattice states in the finite set magnetic oscillator basis of the form $(\emptyset_0, \emptyset_1, \emptyset_2 \dots, \emptyset_{n-1+N_L})$ for $A_n$ sublattice and $(\emptyset_{0,}\emptyset_1, \emptyset_2 \dots, \emptyset_{n+N_L})$ for $B_n$ sublattice. For our computation, we choose Landau level cut off index $N_L = 250$ which is sufficient for numerical convergence.

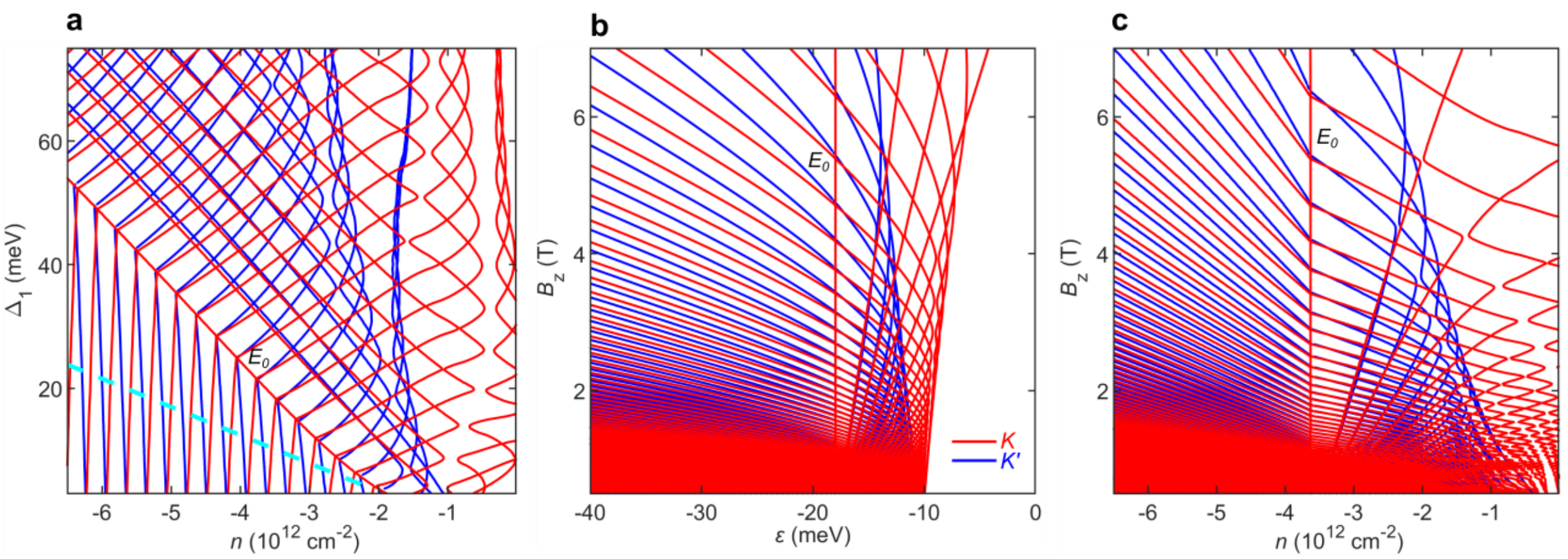

**Extended Data Fig. 3. Calculated Landau levels in the valence band. a,** Evolution of the LLs with $n$ and $\Delta_1$ in valleys $K$ (red) and $K'$ (blue) at $B_z = 3$ T, attained within single particle band structure calculations. The TMBR gives rise to valley polarization and corresponding degeneracy lifting of the LLs. At low carrier densities a full polarization occurs with only $K$ LLs present (red). The red line labeled $E_0$ is the 0$^{th}$ $K$ LL that resides at $k = 0$ and coincides with the annulus opening line $n_a(D)$. The cyan dashed line is the $n_\pi(D)$ line, which traces the positions where the $K$ and $K'$ LLs acquire a relative shift of half a period. These calculated $n_a(D)$ and $n_\pi(D)$ lines are plotted in Figs. 1d-f. **b,** Evolution of the LLs with $\varepsilon$ and $B_z$ for $\Delta_1 =$ 20 meV. **c,** Evolution of the LLs with $n$ and $B_z$ for $\Delta_1 =$ 20 meV.

The LLs have a very interesting evolution as can be seen in Extended Data Fig. 3b. At energies $\varepsilon < E_0$ the Fermi surface is simple and the LLs have a negative slope as expected for hole-like carriers. The 0$^{th}$ $K$ valley LL resides at the bottom of the inverted Mexican hat at $k = 0$. As such, its carriers have no momentum and hence no magnetic moment and therefore its energy $E_0$ is field independent (vertical red line). At $\varepsilon >$ $E_0$ the band structure has an annulus (Fig. 1). The LLs on the inner Fermi surface are electron-like with a positive slope $E_i(B_z)$, while the LLs on the outer Fermi surface have a negative slope. With increasing $B_z$, the electron-like LLs flow towards higher energies, reach the top of the inverted Mexican hat, and then flow towards lower energies along the outer hole-like Fermi surface. Due to the TMBR, the peak energy of the $K$ ($K'$) valence band increases (decreases) with $B_z$. As a result, the turning points of the red (blue) LLs in Extended Data Fig. 3b move up (down) with $B_z$, resulting in a region of fully valley polarized LLs. Similar behavior is observed in Extended Data Figs. 3a,c, showing the evolution of LLs with $n$, $\Delta_1$, and $B_z$, which are the experimentally accessible variables. Note that the TMBR has an effect even in the simple Fermi surface region at $\varepsilon < E_0$, shifting the $K$ and $K'$ LLs and giving rise to the $\pi$-shift line.

Further, to compare with the experimental data, we calculated the DOS in the presence of LLs

$$DOS(\varepsilon) = 2\frac{eB}{h}\sum\nolimits_i \frac{1}{\pi}\frac{\Gamma}{(\varepsilon - E_i)^2 + \Gamma^2}$$

with broadening parameter $\Gamma = 0.8$ meV, chosen to reproduce qualitatively the magnitude of the observed QOs. The prefactor 2 accounts for the spin degeneracy. Figure 1f shows the resulting $DOS(n, D)$ for $B_a = 3$ T in good accord with the experimental data in the full metal state in regions A and B in Fig. 1. In particular, the $\pi$- shift behavior is well reproduced with no fitting parameters. Figure 1f originates from Extended Data Fig. 3a with broadened DOS of the LLs.

**Transport measurements and quantum oscillations**

Magneto transport measurements on the different devices were carried out in dilution refrigerators down to base temperature of $T = 20$ mK by employing standard lock-in techniques at a frequency of 19.18 Hz. By applying $V_{tg}$ and $V_{bg}$ to top and bottom gates, we tune the displacement field $D = (C_t V_{tg} - C_b V_{bg})/2\varepsilon_0$ and the carrier density $n = (C_t V_{tg} + C_b V_{bg})/e$ independently, where $C_t$, $C_b$ are the geometric capacitances per unit area of the top and bottom gates, $\varepsilon_0$ is the vacuum permittivity, and $e$ is the elementary charge.

To analyze the Fermi surface topology, we measured the QOs in $R_{xx}$ vs. $n$ and $B_z$ at various fixed $D$ in device A (Extended Data Fig. 4). As described above, there are two main types of QOs, which arise from the hole-like LLs with negative $B_z$ vs. $n$ slope that are associated with the outer Fermi surface, and the electron-like LLs with a positive slope residing along the inner annular part of the Fermi surface.

We focus first on Extended Data Fig. 4e showing the QOs in $R_{xx}$ at $D = 0.4$ V/nm. At high carrier densities $|n| \gtrsim 3.3\times10^{12}$ cm$^{-2}$ the QOs are well described by the single particle LL calculations presented in Extended Data Fig. 4d. For $|n| \gtrsim 5.4\times10^{12}$ cm$^{-2}$, the Fermi surface is simple for which the QOs vs. $1/B_z$ are described by periodic oscillations of the form $\sin(\frac{2\pi f}{B_z})$, with $f = \frac{nh}{Ne}$. Thus, a four-fold degenerate state ($N = 4$) should display a normalized frequency $f/n = 0.25h/e$, as indeed observed in the FFT spectrum in Extended Data Fig. 4f. For $5.4\times10^{12} \gtrsim |n| \gtrsim 3.3\times10^{12}$ cm$^{-2}$, the Fermi surface is annular which results in two frequencies with the difference between the high frequency due to hole LLs and the low frequency originating from the electron LLs equal to $0.25h/e$. On further decreasing $|n|$, $f/n$ shows two frequencies, satisfying $f^{(1)}/n + f^{(2)}/n \cong 0.5$. This is consistent with partially spin polarized (PSP) phase with two sets of unequally occupied Fermi surfaces. As $|n|$ approaches $2.4\times10^{12}$, $f^{(2)}/n$ vanishes, and $f^{(1)}/n$ converges to $0.5h/e$, implying that the PSP-M phase transforms into a s-1/2M phase where carriers occupy two-fold degenerate simple Fermi surface ($N = 2$). Beyond s-1/2M down to density $|n| \cong 0.9\times10^{12}$ cm$^{-2}$, the normalized QOs frequency is less well defined. However, in this intermediate $|n|$ range, LLs show similar characteristics to the partially polarized phase observed between a-M and s-1/2M for high $B_z$, as shown in Fig. 1e. For $|n|$ between 0.9 to $0.74\times10^{12}$ cm$^{-2}$, Fermi surface becomes simple with degeneracy $N = 1$ (s-1/4M). This results in a normalized frequency $f/n = h/e$.

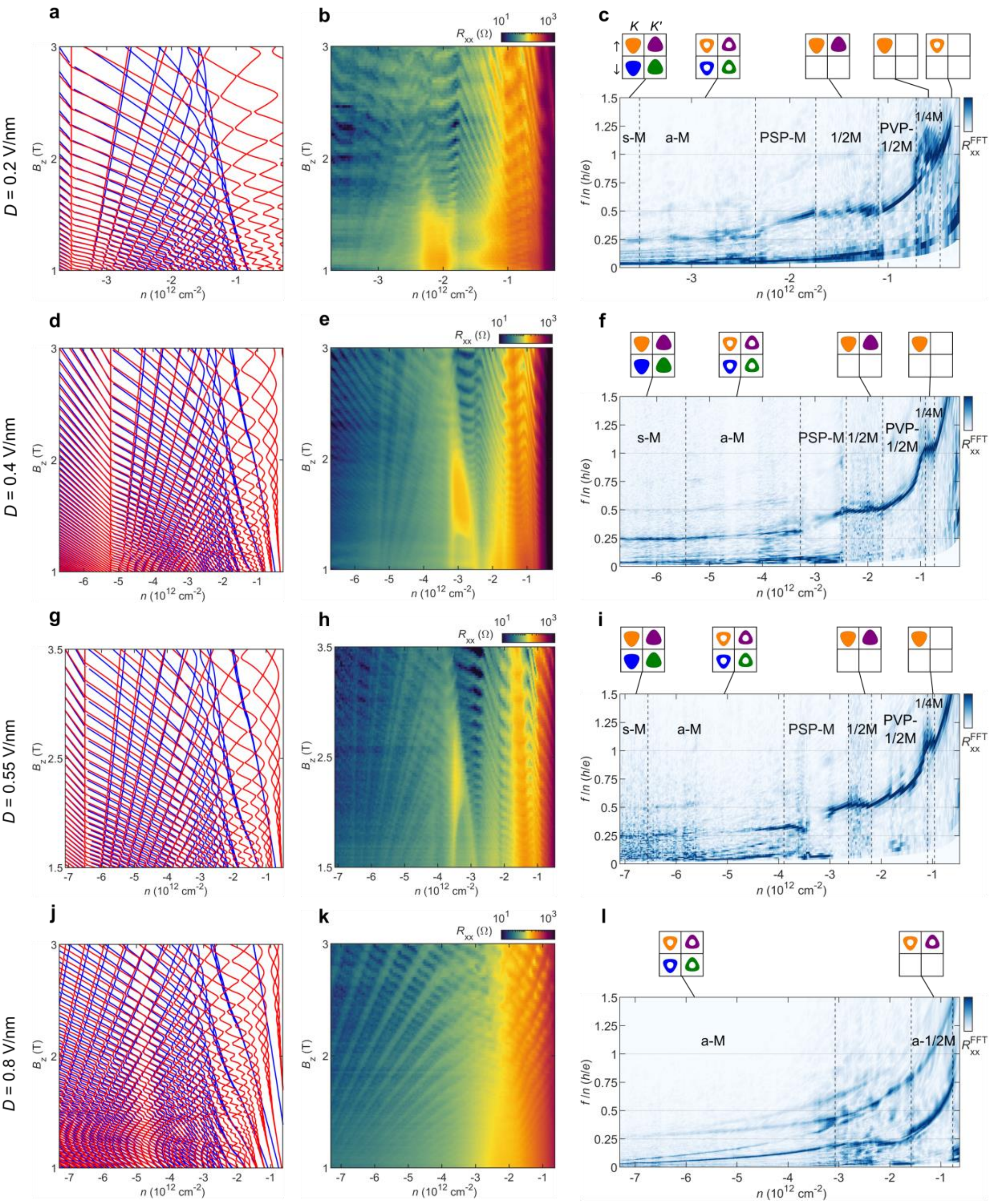

**Extended Data Fig. 4. Analysis of quantum oscillations in $R_{xx}$ in device A. a**, Calculated LLs in $K$ (red) and $K'$ (blue) valleys vs. $n$ and $B_z$ at $D = 0.2$ V/nm within single particle band structure. **b**, Measured $R_{xx}(n, B_z)$ at $D = 0.2$ V/nm. At $|n| \gtrsim 2.4\times10^{12}$ cm$^{-2}$ the QOs are well described by single particle calculations in (a), whereas at lower densities a sequence of interactions-driven symmetry breaking transitions occurs. **c**, Fast Fourier transform (FFT) spectrum of $R_{xx}(1/B_z)$ vs. $n$ with the normalized frequency $f/n$ presented in units of $h/e$. The schematics at the top show the occupied hole states in the four flavors. **d-f**, Same as (a-c) at $D = 0.4$ V/nm. **g-i**, Same as (a-c) at $D = 0.55$ V/nm. **j-l**, Same as (a-c) at $D = 0.8$ V/nm. At high displacement field all the accessible range of carrier densities is described by annular Fermi surfaces.

We observed the same pattern of QOs with varying $n$ for low $D = 0.2$ V/nm and other intermediate $D = 0.55$ V/nm as displayed in Extended Data Figs. 4a-c and 4g-i, respectively. However, at very high $D = 0.8$ V/nm (Extended Data Figs. 4j-l), Fermi surface remains annular over the entire accessible range of $n$, resulting in two distinct peaks in $f/n$. The difference between the peaks remains $0.25h/e$ for $|n| \gtrsim 3.1\times10^{12}$, suggesting that carriers occupied non-interacting four-fold degenerate annular Fermi surface. Below this threshold density, interactions dominate and start lifting the native degeneracies. For $1.6\times10^{12} \gtrsim |n| \gtrsim 0.77\times10^{12}$ cm$^{-2}$, the difference between the high and low frequencies in QOs is $0.5h/e$, implying that the degeneracy of annular Fermi surfaces reduces from four to two.

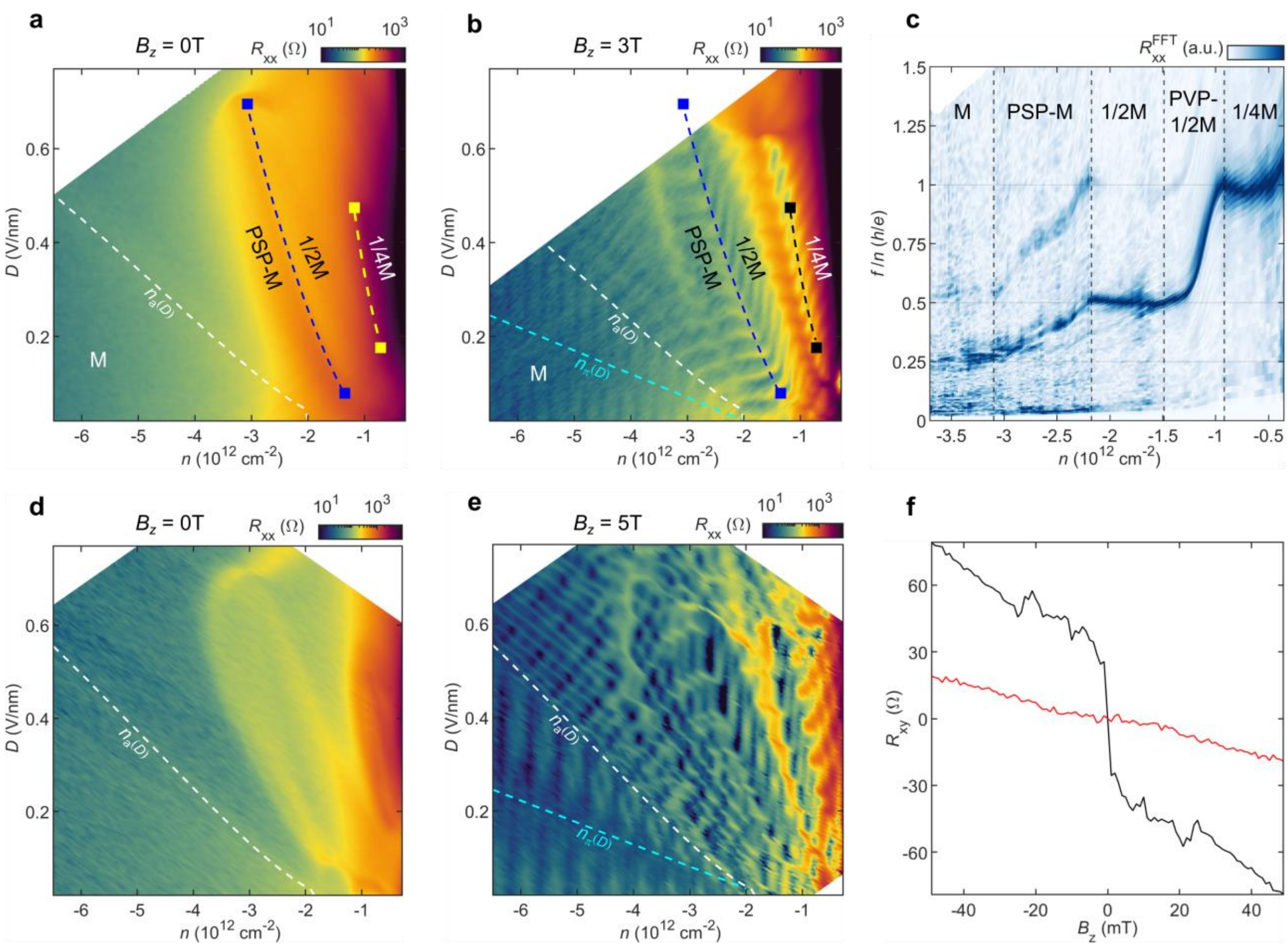

**Extended Data Fig. 5. Transport properties of devices B and C. a,** $R_{xx}(n, D)$ of device B measured at $B_z = 0$ T. The white dashed curve is the $n_a(D)$ line derived from single particle calculations. Blue and yellow lines are M-to-1/2M and 1/2M-to-1/4M transition lines obtained from local magnetization measurements at low field in Fig. 2a. **b**, Same as (a) at $B_z = 3$ T. The cyan dashed curve is $n_\pi(D)$ line derived from single particle calculations. The blue local magnetization transition line closely follows the PSP-M-to-1/2M transition observed in transport. The 1/4M phase at 3 T is seen to expand to slightly higher $|n|$ and significantly larger $D$ as compared to the low-field 1/2M-to-1/4M transition line (black dashed curve). **c**, FFT of the QOs of $R_{xx}(1/B_z)$ vs. $n$ measured at $D = 0.33$ V/nm. The FFT frequency is normalized with respect to Hall carrier density $n_{Hall}$. **d**,**e**, $R_{xx}(n, D)$ of device C measured at $B_z = 0$ (**d**) and 5 T (**e**). **f**, Low-field Hall resistance $R_{xy}$ vs. $B_z$ in device B measured in the 1/2M phase (red line) and in the 1/4M (black line) showing AHE.

Devices B and C showed transport properties and QOs similar to device A, albeit with slightly less pronounced features. Extended Data Figs. 5a,d present $R_{xx}(n, D)$ at $B_a = 0$ T in the two devices, which compares well with Fig. 1d. Also, the QOs in Extended Data Fig. 5b closely follow those in Fig. 1e. In addition, we overlay in Extended Data Figs. 5a,b the magnetic transition lines attained in Fig. 2a marking the M-to-1/2M (blue) and 1/2M-to-1/4M (yellow or black) phase boundaries at low magnetic fields. At $B_a = 0$ T the magnetic transition lines overlap the corresponding distinct features in $R_{xx}(n, D)$. At $B_a = 3$ T (Extended Data Fig. 5b), the blue magnetic transition line overlays the PSP-M-to-1/2M boundary. In contrast, the 1/4M phase expands with $B_a$ to slightly higher $|n|$ and to significantly higher $D$ values (see Extended Data Fig. 6) as compared to zero field (black dashed line).

A clear anomalous Hall effect (AHE) at low fields is observed in the 1/4M phase (black curve in Extended Data Fig. 5f), consistent with spontaneous valley polarization, similar to observations in other rhombohedral structures [15,30]. In contrast, the spin polarized 1/2M phase shows no AHE (red curve) because of the opposite Berry curvature in the two degenerate valleys. These findings are consistent with the local magnetization measurements.

**Expansion of 1/4M state with $B_z$**

The 1/4M domain expands with increasing $B_z$ both in $n$ and $D$. The expansion towards larger $|n|$ is quite small, but resolvable in Extended Data Figs. 5b and 6c. The expansion in $D$, however is very large as seen in Extended Data Figs. 5b and 6a,b. This distinctive expansion is the result of competition between the electric polarization $P$ and the orbital magnetization $\mathcal{M}$. The position of the phase boundary between the 1/4M and the 1/2M above it at higher $D$, is determined by the balance between the change in the electric polarization energy ($PdD$) and the change in the magnetization energy ($\mathcal{M}dB$) at finite $D$ and $B_z$. For the same $n$, the 1/4M possesses less electric polarization due to larger Fermi surface compared to 1/2M. Consequently, the expansion of 1/4M into the 1/2M region with $B_z$ reduces the electric polarization energy gain but enhances the magnetic energy gain, which overall minimizes the total system energy and favors more magnetically polarized state. In Extended Data Fig. 6d, we plot $D_{top}$, corresponding to the top endpoint of the 1/4M phase boundary, as function of $B_z$, and find a linear dependence with a slope $dD_{top}/dB_z = 0.09$ V/nm·T for $B_z < 1.4$ T.

**Self-consistent Hartree-Fock model**

We performed a momentum-resolved mean field analysis of the model Hamiltonian $H = H_0 + H_C + H_{Hu}$, where $H_0$ is the single particle contribution defined above, $H_C$ accounts for the dual-gate-screened Coulomb interaction $V_q = \frac{e^2}{2\epsilon_0}\tanh(qd)/(\epsilon q)$, and $H_{Hu}$ constitutes the short-ranged ferromagnetic Hund's coupling with strength $J_H$. The Hamiltonian $H$ captures the following key properties of the system: The long-ranged Coulomb interaction couples only density-density within each valley, which means it is entirely symmetric in isospin space. In absence of the Hund's coupling, all symmetry-broken states therefore break the isospin symmetry spontaneously. The Hund's term, which is weak compared to the Coulomb interaction, slightly breaks the degeneracy between valley and spin polarization and is therefore important in distinguishing between a spin-polarized and valley-polarized 1/2M, which would otherwise be degenerate. Thus, $H$ presents a minimal model for the experimentally observed sequences of phase transitions.

Based on the experimental findings, we chose $J_H = 3.1$ meV·$10^{-12}$ cm$^2$ for the scHF calculations. The single-particle parameters are given in the table above, $d = 18$ nm is the distance to the metallic gates, and we tune the dielectric constant to attain the best overall fit to the experimental phase diagram with optimal $\epsilon = 23$. Even though the dielectric constant of hBN is much smaller than this value, within the scHF calculation, $\epsilon$ presents an adjustable tuning parameter which encapsulates both external screening and screening within the RMG. As the actual screening strength depends on the density, the fitted value of $\epsilon$ should be viewed as an estimate of the average impact of screening. For the scHF calculation, we project $H$ into the valence band, and search for the minimal free energy in the subspace spanned by both spin and valley degrees of freedom. We explicitly search for flavor degeneracy breaking and intervalley coherent (off-diagonal) order, while taking into account the momentum-dependent renormalization of the dispersion. For the simulation, we use a grid size of 61×61 and convergence is achieved when the energy changes by less than $\Delta E = 10^{-9}$meV between iterations. Further details of this calculation are documented in [27].

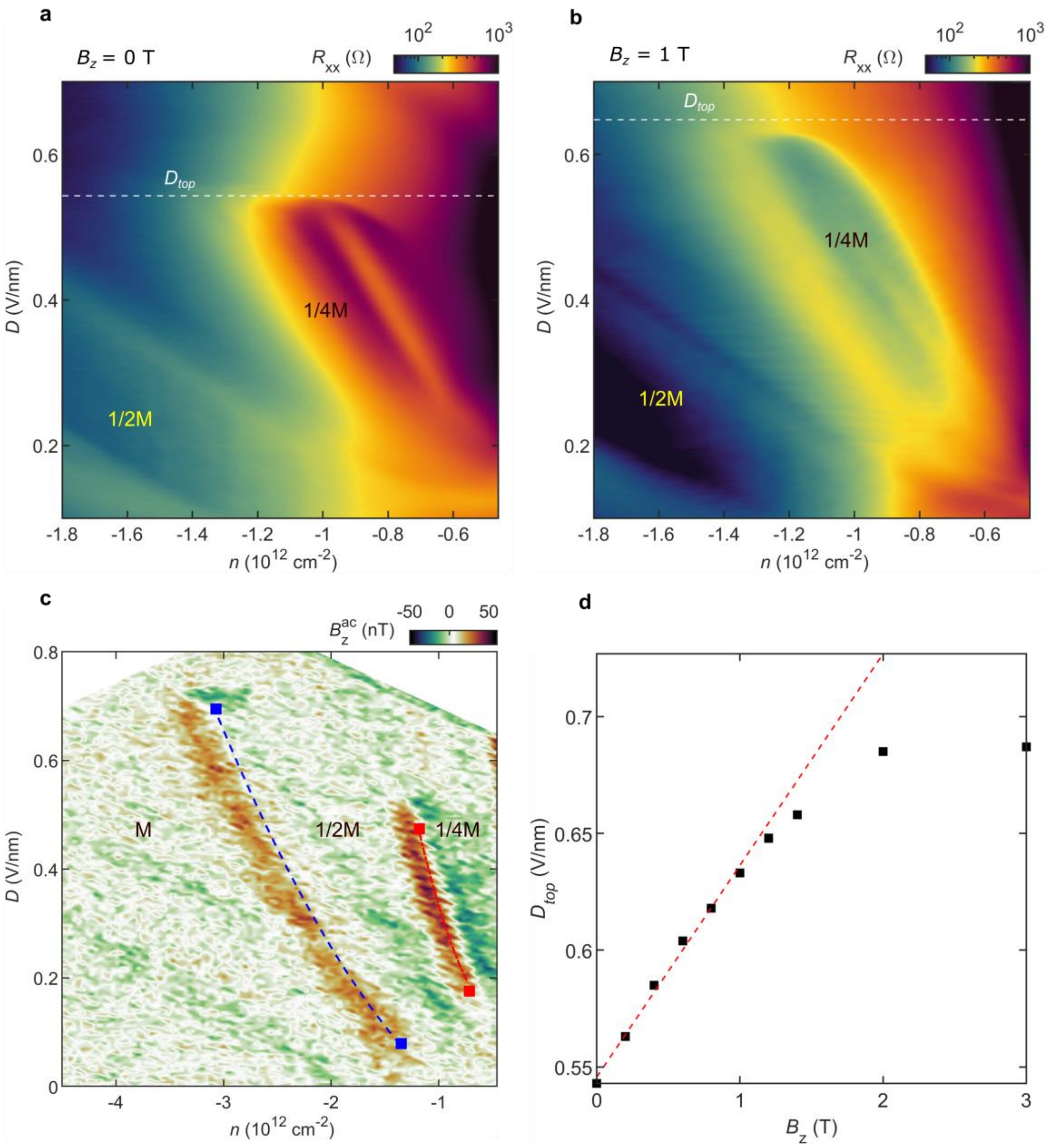

**Extended Data Fig. 6. Expansion of the 1/4M phase with magnetic field. a,** Measured $R_{xx}(n, D)$ in device A at $B_z = 0$ T. The dashed line marks the top endpoint $D_{top}$ of the 1/4M phase. **b,** Same as (a) at $B_z = 1$ T. **c,** $B_z^{ac}(n, D)$ measured in device B at $B_z = 356$ mT. The blue and red curves show the locations of the M-to-1/2M and 1/2M-to-1/4M transition lines at $B_z = 10$ mT derived from Fig. 2b. **d,** Dependence of $D_{top}$ on $B_z$, showing the upward expansion of the 1/4M phase on expense of 1/2M.

The resulting phase diagram is shown in Fig. 2c, providing an essential insight into the fermiology of the various phases and phase transitions. It provides a good general agreement with the experimental diagram in Fig. 2b, with all the experimentally resolved phases discerned in scHF diagram. The same single-particle hopping parameters and screened Coulomb interaction were used in scHF calculations for ABCA and ABC graphene [27], indicating that the successful match between experiment and theory is not a result of fine-tuning. Interestingly, there are two noticeable differences: the a-M with annular Fermi which is observed over an extensive range of parameters in experiment, is found only at high $D$ in the numerics, and at intermediate $n$ and $D$, the scHF calculations give rise to a three-quarter metal (3/4M) that should be accompanied by magnetism and anomalous Hall, which is not observed experimentally. This difference can be the result of the scHF mean-field approximation. Similar disparities were seen in calculations for aluminum arsenide which is another multivalley semi-metal where scHF predicts a 3/4 phase, while variational Monte-Carlo [58] finds that such phase does not exist, in agreement with experiments. Note that in contrast to trilayer RMG [27], no tendency to IVC order in tetralayer RMG is found in scHF simulations. This is likely due to the three pockets at the lowest densities being much less pronounced in the single-particle dispersion for four layers [12], which reduces the amount of kinetic energy gained by forming an IVC.

**Isospin anisotropy analysis and derivation of spin-orbit and Hund's coupling bounds**

To understand the interplay between valley polarization and spin-moment anisotropy, we use a mean-field-inspired model assuming the isospin polarization is momentum independent, which allows us to parametrize the system as a density matrix, using two spinors (assuming no intervalley coherence)

$$\hat{\rho} = \frac{n_K}{n}|\Psi_{\mathrm{k}}\rangle\langle\Psi_{\mathrm{k}}| + \frac{n_{K'}}{n}|\Psi_{\mathrm{k'}}\rangle\langle\Psi_{\mathrm{k'}}|$$
$$|\Psi_{\mathrm{k}}\rangle = a_1|\uparrow K\rangle + a_2|\downarrow K\rangle, \qquad |\Psi_{\mathrm{k'}}\rangle = a_3|\uparrow K'\rangle + a_4|\downarrow K'\rangle, \tag{S1}$$

Where $n = n_K + n_{K'}$ and $|\Psi_{\mathrm{k}}\rangle, |\Psi_{\mathrm{k'}}\rangle$ are both normalized to 1. The energy per carrier is given by

$$E = \frac{1}{2}\chi_\tau^{-1}\langle\tau_z\rangle^2 - U_{Hu}\langle\bar{S}_+\rangle\cdot\langle\bar{S}_-\rangle - \frac{\lambda_{SOC}}{2}\langle\tau_z S_z\rangle - \frac{g}{2}\mu_B\overline{B}\cdot\langle\bar{S}\rangle - \mu_v B_z\langle\tau_z\rangle, \tag{S2}$$

where $\langle\bar{S}_\pm\rangle \equiv \frac{1}{2}\langle(1\pm\tau_z)\bar{S}\rangle$ are valley-spin polarization vectors, $\chi_\tau$ is the valley susceptibility, $\mu_v$ is the valley orbital magnetic moment, the spin and valley operators are taken to have eigenvalues $\pm 1$ and we take $g = 2$ for the $g$ - factor. The interaction energy $U_{Hu} = n * J_H$ is the Hund's coupling energy scale, determined by the carrier density times the intrinsic Hund's coupling, and we assume it is ferromagnetic, as there are plenty of evidence that the 1/2M phase is a spin ferromagnet. The Ising SOC ($\frac{\lambda_{SOC}}{2}\langle\tau_z S_z\rangle$) can be related to a Kane-Mele type SOC in monolayer graphene. In monolayer graphene, the Ising SOC vanishes due to either inversion or $C_2$ rotation symmetry and does not result in any observable spin anisotropy. However, as rhombohedral graphene under displacement field breaks the inversion symmetry and does not have $C_2$ rotation symmetry to begin with, it develops a finite sublattice polarization, and the Kane-Mele type SOC of the graphene monolayer translates to an Ising type SOC with magnitude proportional to the amount of sublattice polarization of the Bloch wavefunctions. In general, for a fixed displacement field, the amount of sublattice polarization will increase with the number of layers, as it is related to the potential difference between the top and the bottom layer.

By using the following parametrization:

$$\hat{\rho} = \frac{1+\delta n}{2}|\Psi_{\mathrm{k}}\rangle\langle\Psi_{\mathrm{k}}| + \frac{1-\delta n}{2}|\Psi_{\mathrm{k'}}\rangle\langle\Psi_{\mathrm{k'}}|$$
$$|\Psi_{\mathrm{K}}\rangle = (\cos\theta_-\,|\uparrow\rangle + \sin\theta_-\,|\downarrow\rangle)\otimes|K\rangle \qquad |\Psi_{\mathrm{K'}}\rangle = (\cos\theta_+\,|\uparrow\rangle + \sin\theta_+\,|\downarrow\rangle)\otimes|K'\rangle, \tag{S3}$$

where $\theta_\pm = \theta_s \pm \delta\theta$, we find

$$E = \frac{1}{2}\chi_\tau^{-1}(\delta n - \chi_\tau\mu_v B_z)^2 - U_{Hu}\frac{1-\delta n^2}{4}\cos(2\delta\theta) - \frac{\lambda_{SOC}}{2}(\sin\theta_s\sin\delta\theta + \delta n\cos\theta_s\cos\delta\theta) -$$
$$-\frac{g}{2}\mu_B[B_x(\sin\theta_s\cos\delta\theta - \delta n\cos\theta_s\sin\delta\theta) + B_z(\cos\theta_s\cos\delta\theta + \delta n\sin\theta_s\sin\delta\theta)]. \tag{S4}$$

Physically, $\theta_s$ is the average spin angle, $\delta\theta$ is the relative spin-canting angle between the valleys, and $\delta n$ is the relative valley density imbalance normalized by the total density as illustrated in Extended Data Fig. 7.

Assuming $\lambda_{SOC}$ to be a small energy scale relative to $\chi_\tau^{-1}$ and $U_{Hu}$, we find the ground state to be a 1/4M ($\langle\tau_z\rangle = \pm 1$) for $\chi_\tau^{-1} + \frac{|U_{Hu}|}{2} < 0$ and a 1/2M ($\langle\tau_z\rangle = 0$) otherwise.

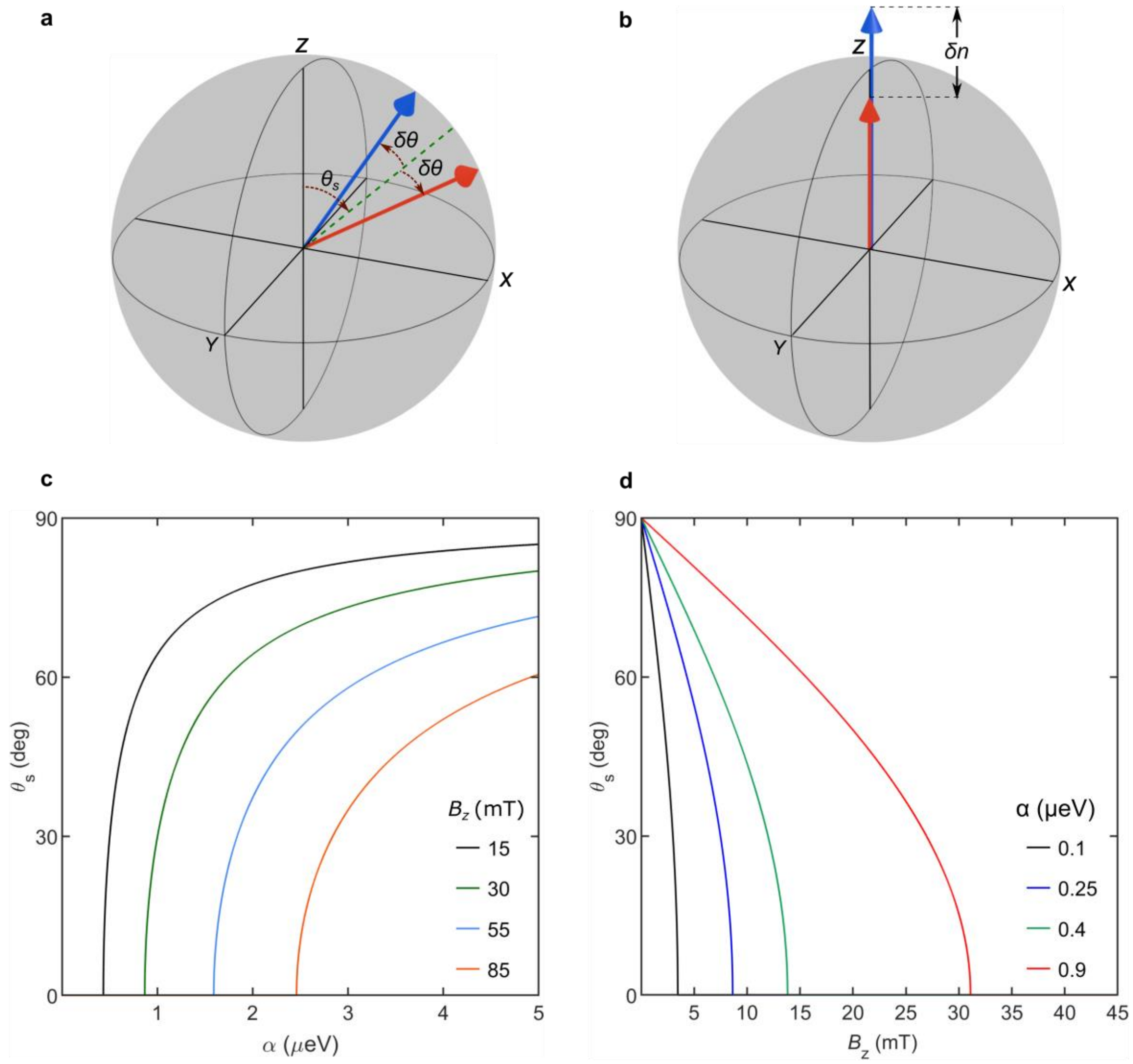

**Extended Data Fig. 7. Spin orientation in 1/2M. a,b**, Schematics of valley-spin polarization vectors $\langle\bar{S}_+\rangle$ (blue) and $\langle\bar{S}_-\rangle$ (red) for the easy-plane (**a**) and easy-axis (**b**) ferromagnets in the presence of small SOC and finite $B_z$. **c**, Spin orientation $\theta_s$ vs. anisotropy parameter $\alpha \geq 0$ (easy-plane) calculated from Eq. S7 in 1/2M for $B_x = 0$ and various indicated values of $B_z$. The black line shows that for the experimentally used $B_z = 15$ mT the spin remains along $\hat{z}$ for anisotropy $\alpha$ below a critical $\alpha_c = 0.43$ µeV, above which the spin unlocks abruptly from $\hat{z}$ orientation and rapidly rotates towards $\hat{x}$ axis ($\theta_s = 90°$). The data in Fig. 4f show that for $\theta_B = 0°$ with $B_z = 15$ mT and $B_x = 0$, the measured magnetization is along $\hat{z}$, which means that $\alpha \leq \alpha_c$. For larger $B_z$ (green, blue, and red curves) the spin remains locked to $\hat{z}$ up to higher values of $\alpha_c$. **d**, Spin orientation $\theta_s$ vs. $B_z$ for $B_x = 0$ and various anisotropy values $\alpha$. For small $\alpha$, the $B_z$ rapidly rotates the spin from the easy plane orientation ($\theta_s = 90°$) towards the $\hat{z}$ axis ($\theta_s = 0°$). With increasing anisotropy, the in-plane magnetization survives up to higher $B_z$ (colored lines).

Quarter metal

As valley polarization breaks TRS, in the 1/4M phase there is an effective spontaneous magnetic field acting on the spin degree of freedom, originating from the SOC. We assume full valley polarization $\delta n = \pm 1$, and find

$$E_{QM} = -\frac{\lambda_{SOC}}{2}\langle \tau_z S_z\rangle - \frac{g}{2}\mu_B\, \bar{B}\cdot\langle\bar{S}\rangle = \mp\frac{\lambda_{SOC}}{2}\langle S_z\rangle - \frac{g}{2}\mu_B\, \bar{B}\cdot\langle\bar{S}\rangle\,. \tag{S5}$$

The sign of the valley polarization $\langle\tau_z\rangle = \delta n$, can be trained by an external field.

We derive the angle of the spin moment from the z-axis for a positive value of $B_z$:

$$\tan\theta_s = \frac{B_x}{B_z + \frac{\lambda_{SOC}}{g\mu_B}}\,. \tag{S6}$$

Figure 4f shows the calculated $\theta_s$ vs. $\theta_B$ from Eq. S6 for fixed $B_z = 15$ mT and varying $B_x$ for two values of $\lambda_{SOC}$ in the 1/4M state. From comparison to the 1/4M data, we can set a lower bound on SOC of $\lambda_{SOC} \gtrsim$ 120 µeV.

Half metal

In the 1/2M, the total magnetic moment includes both valley and spin moments. It is defined as

$$\overline{\mathcal{M}} = -\bar{\nabla}_B E_0,$$

where $E_0$ is the ground state energy

$$E_0 = \min\left\{E(\theta_s,\delta\theta,\delta n) \mid \theta_s\in[0,\pi], \delta\theta\in\left[-\frac{\pi}{2},\frac{\pi}{2}\right], \delta n\in[-1,1]\right\}.$$

We expect both $\delta\theta$ and $\delta n$ to be perturbatively small in $\lambda_{SOC}/U_{Hu}$, as indeed confirmed numerically by minimizing the energy $E$ in Eq. S4. Thus, the energy of the 1/2M can be expressed as a function of $\theta_s$ as the only variable. We do this by minimizing the expression for $E$ over $\delta\theta$ and $\delta n$, and find

$$E_{eff} \approx \alpha\langle S_z\rangle^2 - \frac{g}{2}\mu_B\bar{B}\cdot\langle\bar{S}\rangle = \alpha\cos^2(\theta_s) - \frac{g}{2}\mu_B B\cos(\theta_s-\theta_B), \tag{S7}$$

where

$$\alpha = \left(\frac{\lambda_{SOC}}{2}\right)^2\left(\frac{1}{2U_{Hu}} - \frac{1}{U_{Hu}+2\chi_\tau^{-1}}\right) = \frac{\lambda_{SOC}^2}{8U_{Hu}}\left(\frac{2\chi_\tau^{-1}-U_{Hu}}{2\chi_\tau^{-1}+U_{Hu}}\right) \underset{\chi_\tau^{-1}\gg U_{Hu}}{\approx} \frac{\lambda_{SOC}^2}{8U_{Hu}}\,, \tag{S8}$$

with easy-plane ferromagnet for $\alpha > 0$, and an easy-axis ferromagnet otherwise. Intuitively, the first term of the expression for $\alpha$ comes from the fact that easy-plane ferromagnet gains energy from SOC by canting the spin in each valley in opposite direction (towards $\langle\tau_z\rangle\cdot\hat{z}$), which comes at the cost of intervalley exchange energy, because the spin moments of the two valleys are not perfectly aligned. The second term is the energy gained by an easy-axis ferromagnet due to spin-orbit coupling. In this case, the system gains energy by generating a finite valley polarization ($\langle\tau_z\rangle \neq 0$). From scHF numerical calculations we derive $\chi_\tau = 15.8$ eV$^{-1}$. Since $\chi_\tau^{-1} = 63.3$ meV $\gg U_{Hu}$ the second term is small and hence we expect positive $\alpha$ and correspondingly easy-plane spin ferromagnet.

The 1/2M solid lines in Fig. 4f show the calculated $\theta_s$ vs. $\theta_B$ by minimizing $E_{eff}$ in Eq. S7 for various indicated positive and negative values of $\alpha$ for fixed $B_z = 15$ mT and varying $B_x$. From comparison to the data we can set an upper bound on $|\alpha| < 0.25$ µeV. Combining with the derived $\lambda_{SOC} \gtrsim 120$ µeV and the numerically attained $\chi_\tau^{-1}$, we can set a lower bound on the interaction energy

$$U_{Hu} > 6.5 \text{ meV} \Rightarrow J_H > 3.1 \text{ meV}\cdot 10^{-12}\text{ cm}^2,$$

where we used the experimental carrier density, $n = 2.1\times10^{12}$ cm$^{-2}$. Crucially, this bound is not very sensitive to the numerically extracted value of $\chi_\tau$ as long as $\chi_\tau^{-1} \gg U_{Hu}$ in Eq. S8.

With these parameters, $\delta n$ is vanishingly small and hence the orbital magnetization is negligible. As a result, the magnetization in the 1/2M phase is given by the spin magnetization. Similarly, $\delta\theta$ is very small and hence the spin canting is negligible. Note that the attained value of $U_{Hu} > 6.5$ meV is not much lower than the maximal theoretical value of $U_{Hu} = V_{q=K}(\epsilon = 1) = 11.2$ meV, which is the intervalley Coulomb energy in vacuum ($\bar{K} = (0, \frac{4\pi}{3})$ is the intervalley wavevector). Note that the dielectric constant at wavevector $\bar{K}$ can be significantly different from the dielectric constant at small wavevectors, that enters the screened Coulomb interaction at the distance between electrons.

In the above discussion we have assumed momentum independent Hund's coupling. In the presence of a $k$-dependent coupling, the anisotropy expression $\alpha \cong \lambda_{SOC}^2/8U_{Hu}$ in the limit of $\lambda_{soc} \ll U_{Hu}$ will remain valid, where $U_{Hu}$ should be thought of as momentum-averaged Hund's energy (i.e. it remains the average Hund's interaction energy per electron). Our reasoning for the Hund's origin of spin isotropy in 1/2M is as follows: Given the fact that the 1/4M phase is fully out of plane polarized, we infer that the spin orientation in each valley is dominated by Ising-like SOC term. In this case the only way to get an isotropic non-zero spin magnetism in the 1/2M phase is by coupling the spins between the valleys such that they will have a common orientation and the effect of the SOC on each valley will cancel out. The simplest term that has this effect and is allowed by the symmetries of the problem is an inter-valley Hund's exchange coupling. We are not aware of any additional symmetry-allowed term that under reasonable assumptions can explain the observed spin isotropy in the 1/2M state.

**Experimental determination of isospin tilt angle in 1/2M**

The tilt angle of the isospin moment at M-to-1/2M transition was determined as follows. The magnetic moment distribution $\boldsymbol{m}(\boldsymbol{r}) = m(\boldsymbol{r})\hat{z}$ attained in out-of-plane $B_a$ ($\theta_B = 0$, Extended Data Fig. 7a) is rotated by a trial angle $\theta_{tr}$, $\boldsymbol{m}(\boldsymbol{r}) = m(\boldsymbol{r})\cos\theta_{tr}\,\hat{z} + m(\boldsymbol{r})\sin\theta_{tr}\,\hat{x}$ and the corresponding $B_z^{tr}(\boldsymbol{r}, \theta_{tr})$ is calculated,

$$B_z^{tr}(\boldsymbol{r}, \theta_{tr}) = \frac{\mu_0}{4\pi} \iint \left[ \frac{3\boldsymbol{m}(\boldsymbol{r}') \cdot (\boldsymbol{r} - \boldsymbol{r}')}{|\boldsymbol{r} - \boldsymbol{r}'|^5} (\boldsymbol{r} - \boldsymbol{r}') \cdot \hat{z} - \frac{\boldsymbol{m}(\boldsymbol{r}') \cdot \hat{z}}{|\boldsymbol{r} - \boldsymbol{r}'|^3} \right] dA'.$$

We then calculate the mean squared deviation between the measured $B_z^{ac}(\boldsymbol{r})$ in presence of $B_a$ along a given $\theta_B$ (Extended Data Fig. 8b) and the calculated $B_z^{tr}(\boldsymbol{r}, \theta_{tr})$ (Extended Data Fig. 8c), quantified by $\mathrm{Error} = \langle |B_z^{ac}(\boldsymbol{r}, \theta_B) - B_z^{tr}(\boldsymbol{r}, \theta_{tr})|^2 \rangle$. We identify the isospin tilt angle $\theta_s$ with $\theta_{tr}$ that results in minimal error as shown in Extended Data Fig. 8d. The resulting $\theta_s$ for various positive and negative $\theta_B$ is presented in Fig. 4f.

In the above procedure we have made a simplifying assumption that the spin tilt angle $\theta_s$ is position independent, whereas disorder and local strain may result in position dependent magnetic anisotropy. We note that the inversion problem with arbitrary spin orientation has no single-valued mathematical solution, and hence some constraints must be imposed. We have therefore applied the simplest assumption of uniform $\theta_s$ that shows a good fit to the experimental data.

To provide additional assessment of the applicability of this assumption and to test for spatial variations in $\theta_s$, we have repeated the analysis over several narrow strips of the sample, as marked by the colored dashed rectangles in Extended Data Fig. 8e. To improve the signal to noise ratio, we integrated the $B_z^{ac}(x, y)$ over $y$ within each strip. The red curves in Extended Data Fig. 8f show the averaged $B_z^{ac}(x)$ profiles in the three strips attained from Fig. 3g, as compared to the calculated ones from Extended Data Fig. 8e (blue) for the case of $\theta_B = 0$. Extended Data Fig. 8g shows the results of the same procedure for $\theta_B = -45°$. The corresponding $\mathrm{Error}$ function for the three strips for $\theta_B = -45°$ is shown in Extended Data Fig. 8h. The quality of the fits and the derived uncertainty of the tilt angles $\theta_s$ in the three strips is consistent with global values in Extended Data Fig. 8d within our uncertainty of 5°, suggesting that the variations in the local anisotropy are weak, and the orientation of the magnetization can be considered homogenous.

Our understanding of how strain affects electronic properties is generally limited, and its impact on magnetic anisotropy is even less understood. In the case of rhombohedral tetralayer graphene, we believe that strain's effect on the local magnetization orientation should be weak. This conclusion is based on our finding that the 1/2M is highly isotropic, while the 1/4M is strongly easy-axis anisotropic. The theoretical curves in Fig. 4f show that substantial variations in $\alpha$ in 1/2M and in $\lambda_{SOC}$ in 1/4M, result in relatively small deviations in the resulting spin orientation $\theta_s$. Therefore, we believe that in practice, possible strain-induced local variations in the spin-orbit coupling or anisotropy parameter will result in only small local variations in $\theta_s$, unresolvable within our experimental resolution.

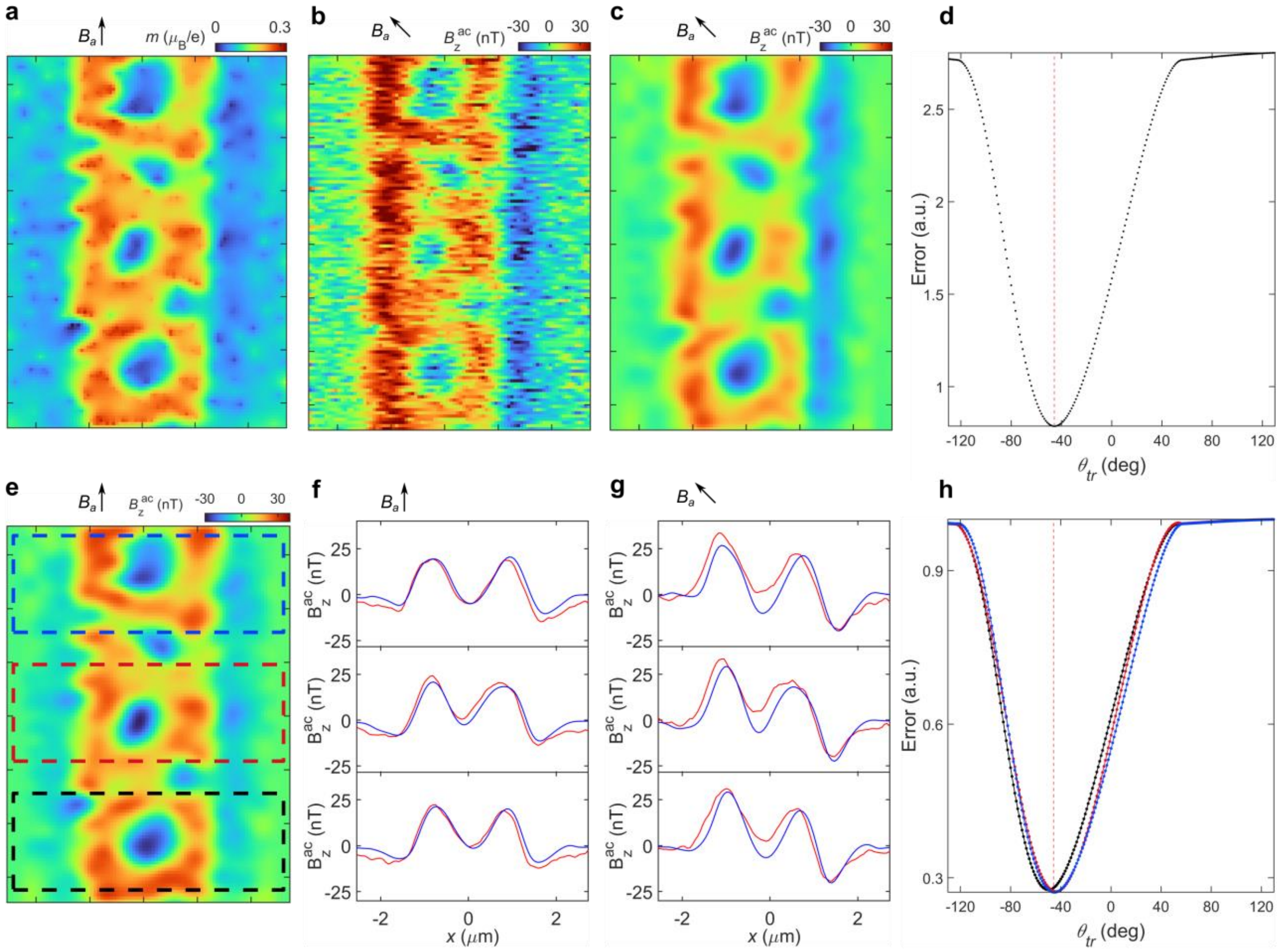

**Extended Data Fig. 8. Determining isospin tilt angle at M-to-1/2M transition. a,** Reconstructed differential magnetization $m(x,y)$ in presence of $B_a = B_z$ ($\theta_B = 0$) reproduced from Fig. 3j. **b,** $B_z^{ac}(x,y)$ measured at M-to-1/2M transition (black star in Fig. 2a) at $\theta_B = -45°$ (Fig. 3i). **c**, Calculated $B_z^{ac}(x,y,\theta_{tr})$ for $\theta_{tr} = -45.6°$, at which the mean squared deviation between (b) and (c) is minimal. **d**, Plot of the mean squared error vs. the trial angle $\theta_{tr}$. The dashed line marks $\theta_{tr} = -45.6°$ with the minimal error, which defines the spin tilt angle $\theta_s$. **e,** Calculated $B_z^{ac}(x,y)$ from $m(x,y)$ in (a) for $\theta_B = 0°$. **f,** Comparison between the measured $B_z^{ac}(x)$ profiles in Fig. 3g (red) and the computed ones in (e) (blue), averaged along $y$ in the three dashed-border rectangular strips marked in (e). **g,** Same as (f) comparing the averaged measured $B_z^{ac}(x)$ from (b) (red) with the calculated one from (c) (blue). **h,** The mean squared error as in (d) computed for the three strips marked in (e). Red dashed line marks $\theta_{tr} = \theta_s = -45.6°$.

## Evaluation of the isospin texture in the 1/4M phase

Extended Data Fig. 9a shows schematically the different components of the magnetization vectors at the 1/2M-to-1/4M transition. The measured differential magnetization across the transition, $\boldsymbol{m}$ (magenta vector), is given by the difference between the magnetization in the 1/4M, $\boldsymbol{\mathcal{M}_{1/4M}}$ (blue vector), and in the 1/2M, $\boldsymbol{\mathcal{M}_{1/2M}}$ (purple): $\boldsymbol{m} = \boldsymbol{\mathcal{M}_{1/4M}} - \boldsymbol{\mathcal{M}_{1/2M}} = \boldsymbol{\mathcal{M}^o_{1/4M}} + \boldsymbol{\mathcal{M}^s_{1/4M}} - \boldsymbol{\mathcal{M}^s_{1/2M}}$. Here $\boldsymbol{\mathcal{M}^s_{1/2M}}$ is the spin magnetization in 1/2M (purple) oriented along the applied field direction $\theta_B$, and $\boldsymbol{\mathcal{M}_{1/4M}}$ (blue) is given by the sum of the orbital magnetization, $\boldsymbol{\mathcal{M}^o_{1/4M}}$ (brown), oriented along $\hat{z}$, and the spin magnetization, $\boldsymbol{\mathcal{M}^s_{1/4M}}$ (green), with a tilt angle $\theta_s$ that we aim to determine as follows.

Since the carrier density at the two sides of the 1/2M-to-1/4M transition is the same, the spin magnetization should have the same magnitude $|\boldsymbol{\mathcal{M}^s_{1/2M}}| = |\boldsymbol{\mathcal{M}^s_{1/4M}}|$ (green and purple vectors are of the same length), and should equal to $|\boldsymbol{\mathcal{M}^s_{1/2M}}|$ at the M-to-1/2M transition (Fig. 3j), rescaled by the relative carrier densities at the two transitions and by the partial polarization in PSP-M region. For out-of-plane $B_a$ ($\theta_B = 0$), $\boldsymbol{\mathcal{M}^s_{1/4M}} = \boldsymbol{\mathcal{M}^s_{1/2M}}$ (green and purple vectors are parallel) and hence $\boldsymbol{m} = \boldsymbol{\mathcal{M}^o_{1/4M}}$. The brown vector is thus known and is given by Fig. 4d. We can then evaluate $\theta_s$ by comparing the measured $B_z^{ac}(x, y)$ corresponding to $\boldsymbol{m}$ (magenta), with $B_z^{ac}(x, y, \theta_{tr})$ derived from summation of the green, brown, and purple vectors, $\boldsymbol{m} = \boldsymbol{\mathcal{M}^o_{1/4M}} + \boldsymbol{\mathcal{M}^s_{1/4M}}(\theta_{tr}) - \boldsymbol{\mathcal{M}^s_{1/2M}}(\theta_B)$, where $\theta_{tr}$ is a trial angle for the spin orientation $\theta_s$ in the 1/4M (green).

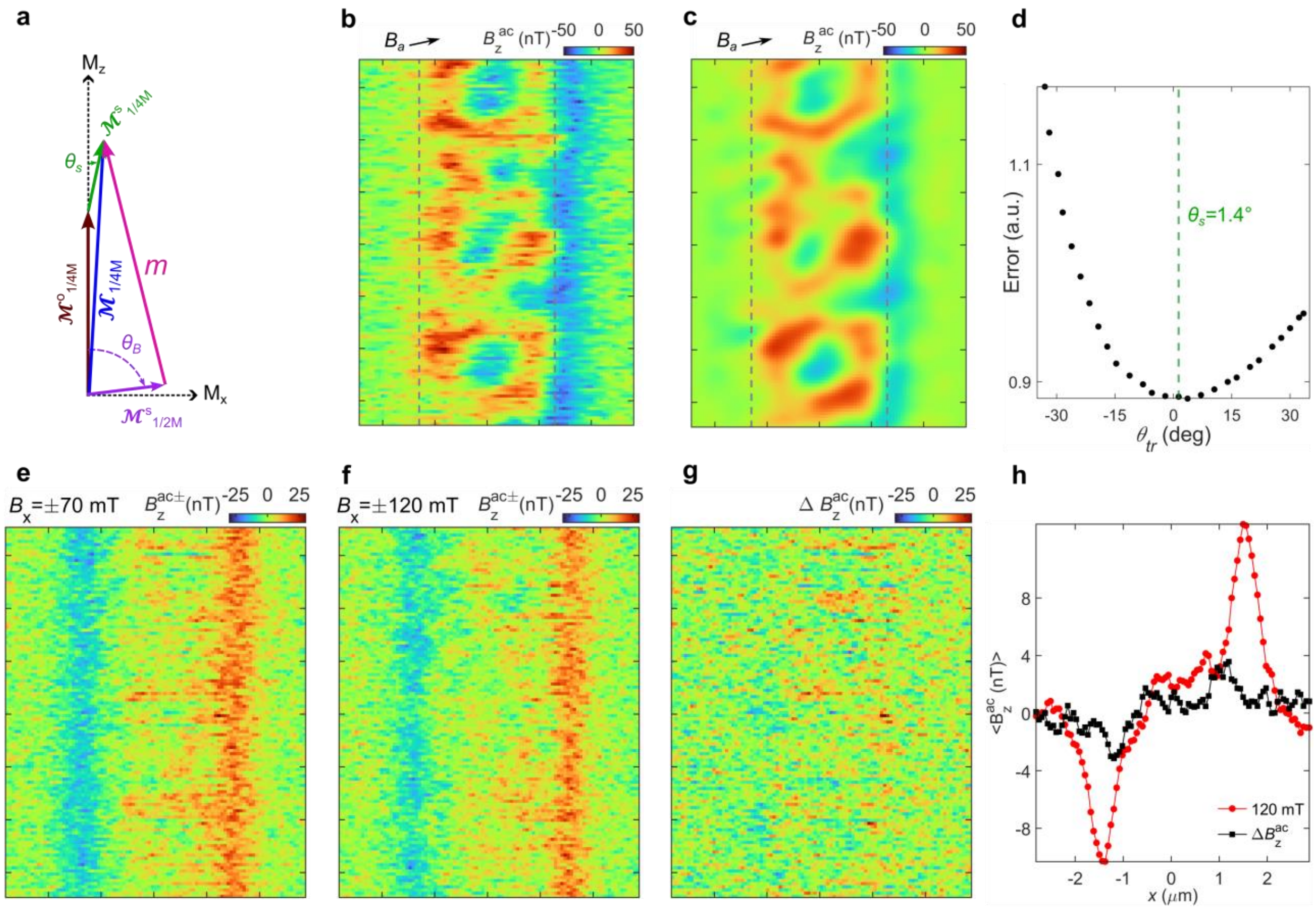

**Extended Data Fig. 9. Analysis of isospin orientation at the 1/2M-to-1/4M transition. a,** Schematic vector representation of isospin moments at 1/2M-1/4M transition in a tilted magnetic field with angle $\theta_B$. **b,** Measured $B_z^{ac}(x, y)$ at 1/2M-1/4M transition for $B_a$ along $\theta_B = 78°$. **c,** Trial $B_z^{ac}(x, y, \theta_{tr})$ calculated from the vector analysis in (a) for $\theta_{tr} = 1.4°$. **d,** Plot of the mean square error vs. the trial angle $\theta_{tr}$. The green dashed line marks $\theta_{tr} = 1.4°$ with the minimal error, which defines the tilt angle $\theta_s$ of $\boldsymbol{\mathcal{M}^s_{1/4M}}$ (green vector in (a)). **e,** $B_z^{ac\pm}(x, y)$ in presence of a small $B_z = 15$ mT, obtained by subtracting $B_z^{ac-}(x, y)$

measured in presence of in-plane magnetic field $B_x = -70$ mT from $B_z^{ac+}(x,y)$ acquired at $B_x = 70$ mT. **f,** Same as (a) at $B_x = 120$ mT. **g**, The difference $\Delta B_z^{ac}(x,y)$ between (a) and (b). **h**, $B_z^{ac\pm}(x)$ from (b) averaged over $y$ axis (red), and $\Delta B_z^{ac}(x)$ from (c) averaged over $y$ (black), allowing estimation of the upper bound of the isospin tilt angle in the 1/4M phase of about 5°.

Extended Data Figs. 9b,c show the measured $B_z^{ac}(x,y)$ at $\theta_B = 78°$ and the calculated $B_z^{tr}(x,y,\theta_{tr})$ using the above procedure, while Fig. 4e shows the calculated $B_z^{tr}(x,y,\theta_{tr})$ at $\theta_B = 83°$. The corresponding mean square error $\mathrm{Error} = \langle |B_z^{ac}(x,y) - B_z^{tr}(x,y,\theta_{tr})|^2 \rangle$ at $\theta_B = 78°$ is plotted as a function of $\theta_{tr}$ in Extended Data Fig. 9d. The minimal error at $\theta_{tr} = 1.4°$ designates the spin tilt angle ($\theta_{tr} = \theta_s$) of $\boldsymbol{\mathcal{M}}_{\mathbf{1/4M}}^{s}$ (green vector) and the resulting $B_z^{ac}(x,y,\theta_{tr} = \theta_s)$ is shown in Extended Data Fig. 9c. The magenta triangles in Fig. 4f show the average $|\theta_s|$ derived from the above procedure at $\theta_B = \pm 78°$ and $\theta_B = \pm 83°$.

We can attain an additional bound on $\theta_s$ in the 1/4M as follows. Because the isospin magnetization is isotropic in the 1/2M, the spin will be essentially in-plane polarized at large in-plane field $B_x$. Further increase in $B_x$ will hardly affect the magnetization in the 1/2M, but may tilt the spin orientation in the 1/4M away from the easy axis. We apply first a small $B_z = 15$ mT and a large $B_x = 70$ mT ($\theta_B = 78°$) and acquire the corresponding $B_z^{ac+}(x,y)$ across the 1/2M-to-1/4M transition. The in-plane field is then set to $B_x = -70$ mT and the resulting $B_z^{ac-}(x,y)$ is measured. The two sets of data are subtracted, with the difference, $B_z^{ac\pm}(x,y) = B_z^{ac+}(x,y) - B_z^{ac-}(x,y)$, presented in Extended Data Fig. 9e. This procedure eliminates the out-of-plane component of the differential magnetization, which is independent of $B_x$ direction, with $B_z^{ac\pm}(x,y)$ reflecting only the stray field due to the in-plane component of the differential magnetization across the transition. The same procedure is repeated with substantially larger $B_x = \pm 120$ mT ($\theta_B = 83°$), with the resulting $B_z^{ac\pm}(x,y)$ presented in Extended Data Fig. 9f. Extended Data Fig. 9g presents the difference $\Delta B_z^{ac}(x,y) = B_z^{ac\pm}(x,y)_{120\,mT} - B_z^{ac\pm}(x,y)_{70\,mT}$, which shows that the change in the differential in-plane magnetization across the transition, and hence the possible change in the isospin orientation in the 1/4M, are unresolvable between the two $B_x$ values within our resolution. By averaging over the $y$ axis for improved signal to noise ratio of the $B_z^{ac\pm}(x,y)_{120\,mT}$ data (red) and the $\Delta B_z^{ac}(x,y)$ data (black) in Extended Data Fig. 9h, we can estimate that the change in the in-plane magnetization is less than 9%, placing an upper bound on possible spin tilt angle change in the 1/4M between $B_x = 70$ and 120 mT of about 5°, which we use as the error bar for $\theta_s$ in the 1/4M phase in Fig. 4f.

**Reconstruction of 2D isospin magnetization**

To reconstruct the out-of-plane magnetization, instead of the Tikhonov regularized Fourier transform-based approach [59], we employ a numerical method based on a deep neural network, similar to [60]. In this method, we consider a three-layer network architecture, consisting of the Input Layer, Deep Layer and recovered Image Layer. Both the Deep Layer and the recovered Image Layer utilize the Rectified Linear Unit activation function. The recovered Image Layer represents the magnetization $m$. The convolution of the recovered Image Layer with the Biot-Savart kernel yields $B_{z,sim}^{ac}$, the magnetic field predicted by the network. To achieve faster convergence, the Input Layer is initialized with the measured values of $B_z^{ac}(x,y)$. The network's weights are optimized by minimizing the merit function,

$$f_M = ||B_{z,sim}^{ac} - B_z^{ac}||^2 + \lambda_\alpha ||\nabla^2 m||^2,$$

which ensures that the theoretical magnetic field closely matches the measured field while maintaining smoothness in the magnetization. Here, $\lambda_\alpha$ is the Tikhonov regularization parameter.